\documentclass[
aps,
 ,twocolumn
]{revtex4}
\usepackage{graphicx}
\usepackage{amsmath}
\usepackage{amssymb}
\usepackage{color}
\usepackage{hyperref}
\hypersetup{colorlinks,linkcolor={blue},citecolor={red},urlcolor={blue}}

\begin{document}

\title{Correcting finite squeezing errors in continuous-variable cluster states}

\author{Daiqin Su}
\email{daiqin@xanadu.ai}
\author{Christian Weedbrook}
\author{Kamil Br\'adler}
\affiliation{Xanadu, 372 Richmond Street West, Toronto, Ontario M5V 1X6, Canada }


\date{\today}

\begin{abstract}
{We introduce an efficient scheme to correct errors due to the finite squeezing effects in continuous-variable cluster states. 
Specifically, we consider the typical situation where the class of algorithms consists of input states that are known. By using the knowledge of the input states, we 
can diagnose exactly what errors have occurred and correct them in the context of temporal continuous-variable cluster states. 
We illustrate the error correction scheme for single-mode and two-mode unitaries implemented by spatial continuous-variable
cluster states. We show that there is no resource advantage to error correcting multimode unitaries implemented by spatial cluster states. 
However, the generalization to multimode unitaries implemented by temporal continuous-variable cluster states shows significant practical advantages 
since it costs only a finite number of optical elements (squeezer, beam splitter, {\rm etc}). }
\end{abstract}

\maketitle

\vspace{10 mm}

\section{Introduction}

Quantum states are fragile and can be easily perturbed by noise and the environment. Many methods have been developed to protect quantum states against noise. 
The most important and widely used method is to develop quantum error correction codes, both for discrete-variable \cite{Shor1995, Steane1996} 
and continuous-variable (CV) quantum states \cite{Braunstein1998, Lloyd1998, Braunstein1998-5wp}.
Typically, the quantum state is encoded in a larger Hilbert space and the quantum 
information is stored within the entanglement between the quantum system and ancillary systems (qubits or qumodes). For CV quantum systems, 
experimental implementations of quantum error correction for single-mode errors have been realized \cite{Aoki2009, Hao2015}.
Quantum error correction methods can also be used in the context of distilling CV entanglement, e.g., by using a noiseless linear amplifier \cite{Ralph2011, Dias2017}. 

Measurement-based quantum computation is one particular model of quantum computing \cite{Raussendorf2001}.
It is based on a resource state called a cluster state \cite{Briegel2001}, where computations are implemented via local measurements on either qubits or qumodes. 
For CV measurement-based quantum computation in terms of Gaussian unitaries, the cluster state is a multimode entangled Gaussian state and the required local measurements are 
homodyne detection and photon counting \cite{Menicucci2006}. The basic building block 
of measurement-based quantum computation is quantum teleportation. By choosing different homodyne measurement quadratures one can implement various unitaries. 
It is known that teleportation is perfect only when the entanglement is maximal, namely, the squeezing is infinite. However, an infinitely squeezed state is unphysical since
it requires infinite energy. Therefore, for a realistic cluster state, the squeezing is finite. This results in imperfections in CV quantum teleportation 
\cite{Pirandola2015} (less than unity fidelity) and thus noise in measurement-based quantum computation \cite{Menicucci2006}. 

The purpose of this work is to correct the errors induced by the finite squeezing in the CV cluster states. Specifically, we consider the case where 
the input state to the quantum algorithm is known. The fact that it is known is the standard case in computation where one has a known input state followed 
by known operations or gates followed by the output. For example, this would include Shor's algorithm \cite{Shor1994}, HHL \cite{Harrow2009}, 
IQP \cite{Bremner2011}. Some exceptions would include quantum teleportation \cite{Pirandola2015} (where the input state is unknown), qPCA \cite{Lloyd2014qpca}, 
and oracle-based schemes such as Grover's algorithm \cite{Grover1996}.

The key idea of this work is to use the full information of the input state to perform error correction. If both the input state and the unitary are known,
the target output state can in principle be calculated. However, due to the effect of finite squeezing the actual output state is different from the target state. 
An additional unitary can be introduced to transform the actual output state to the target state, thus correcting the error. 
It is important that the additional unitaries require finite amounts of optical resources. We show that
in the case of single-mode,  two-mode, as well as multimode unitaries implemented by spatial cluster states, 
there exists no advantage to use this particular error correction scheme. However, in the time domain cluster states practical advantages can 
be obtained as the scaling stays fixed for arbitrary amounts of input modes.


The proposed error correction scheme is different from standard error correction schemes, either based on error correcting codes or the noiseless linear amplifier. 
The standard schemes usually are concerned with interactions between the quantum systems and the environment. While this is also a concern for 
measurement-based CV quantum computation, in this paper we only consider the effect of errors due to finite squeezing effects. Other types of errors will be considered in a later work.
Since both the CV cluster states and the implemented unitaries are Gaussian, the errors are also Gaussian. The proposed error correction scheme thus can use 
Gaussian operations to correct Gaussian errors. This does not violate the no-go theorem \cite{Niset2009} because here there is no encoding involved. 

This paper is organized as follows. In Sec \ref{sec:background}, we briefly discuss some of the basic unitaries, the Wigner function and CV cluster states.  
In Sec. \ref{sec:SingleModeEC}, we illustrate the error correction scheme for single-mode unitaries. Sec. \ref{sec:TwoModeEC} generalizes
the error correction scheme to two-mode unitaries. The single-mode and two-mode cases show the validity of the scheme and we discuss in Sec. \ref{sec:MultiModeEC}
the practical advantage of the scheme when multimode unitaries are implemented by temporal CV cluster states. We conclude in Sec. \ref{sec:Conclusion}. 

\section{Background}\label{sec:background}

\subsection{Basic Gaussian Unitaries}

We consider an $M$-mode quantized optical field that can be described by annihilation operators $\hat a_i$, $i = 1, 2, \cdots, M$. 
They  satisfy commutation relations $[\hat a_j, \hat a^{\dag}_k] = \delta_{jk}$, where $\hat a_k^{\dag}$ is the Hermitian conjugate of $\hat a_k$. 
It is also convenient to describe the optical field using position (amplitude) and momentum (phase) quadratures $\hat q_i$ and $\hat p_i$, 
\begin{eqnarray}
 \hat q_i = \frac{1}{\sqrt{2}} \big(\hat a_i + \hat a^{\dag}_i \big), ~~~~~~  \hat p_i = -\frac{i}{\sqrt{2}} \big(\hat a_i - \hat a^{\dag}_i \big),
\end{eqnarray}
which satisfy $[\hat q_j, \hat p_k]=i \delta_{jk}$ (we set $\hbar = 1$ throughout the paper). 
To obtain compact formulas in the following, we introduce an operator value vector with $2M$ components,
$\hat{\boldsymbol{\xi}} = (\hat q_1, \hat p_1, \cdots, \hat q_M, \hat p_M)^{\top}$. In this paper, we are only concerned with Gaussian states and Gaussian unitaries. 
It is adequate to represent a Gaussian state by writing the first and second moments of the quadratures \cite{Weedbrook2012RMP}. The first moment
is defined as $d_i = \langle \hat \xi_i \rangle$, characterizing the displacement in the quadrature. The second moment is known as the covariance matrix, defined as
\begin{eqnarray}
\sigma_{ij} = \frac{1}{2} \big\langle \big\{ \hat \xi_i, \hat \xi_j^{\dag} \big\}  \big\rangle - d_i d_j,
\end{eqnarray}
where \{ , \} stands for the anti-commutator. In this convention, the covariance matrix of the vacuum state is $\sigma_{\text{vac}} = {\rm diag} \{1/2, 1/2 \}$. 
In the Heisenberg picture, the evolution of the quadratures under a unitary $\hat U$ is 
\begin{eqnarray}\label{eq:unitary}
\hat \xi_i \rightarrow \hat \xi_i^{\,\prime} = \hat U^{\dag} \hat \xi_i  \hat U = \sum_{j=1}^{M} S_{ij} \hat \xi_j + c_i,
\end{eqnarray}
where $S_{ij}$ is called the symplectic matrix and $c_i$ represents the displacement. 
The evolution of the covariance matrix is 
\begin{eqnarray}
\sigma \rightarrow \sigma^{\prime} = S \sigma S^{\top}.
\end{eqnarray}

A general multimode unitary can always be decomposed into a collection of single-mode and two-mode unitaries. It is thus important to introduce some basic
single-mode and two-mode unitaries. 

{\bf Phase shift}: 
The phase shift unitary in the $j$-th mode is 
\begin{eqnarray}
 \hat R_j(\theta) = \exp \bigg[\frac{i \theta}{2} \big( \hat q_j^2 + \hat p_j^2 \big) \bigg].
\end{eqnarray}
The corresponding symplectic matrix $R_j(\theta)$ is 
\begin{eqnarray}
R_j(\theta) = 
 \begin{pmatrix}
 \cos \theta & -\sin \theta \\
 \sin \theta & \,\,\,\,\, \cos \theta 
 \end{pmatrix}. 
\end{eqnarray}

{\bf Single-mode squeezing}: We follow the definition of single-mode squeezing unitary in \cite{Gu2009}, which is different from the standard definition. 
\begin{eqnarray}\label{eq:SMS}
\hat S_j(s) = \exp \bigg[-i \bigg(\frac{\ln s}{2} \bigg) \big( \hat q_j \hat p_j + \hat p_j \hat q_j \big)\bigg],
\end{eqnarray}
where $\ln s$ is the squeezing parameter. By defining it this way, the symplectic matrix takes a simple form,
\begin{eqnarray}
S_j(s) = \begin{pmatrix}
 s & 0 \\
 0 & 1/s 
 \end{pmatrix}.
\end{eqnarray}
It can be seen that $s$ plays the role as a squeezing factor to the position quadrature. 

{\bf Controlled-Z gate}: This is a two-mode unitary, which is defined as
\begin{eqnarray}\label{eq:CZ}
\hat C_Z(g) = \exp \big( i g \,\hat q_j \otimes \hat q_k \big),  ~~~~~~ j \ne k, 
\end{eqnarray}
where $g$ is the interaction strength. Its symplectic matrix is 
\begin{eqnarray}\label{eq:CZSM}
C_Z (g) =
 \begin{pmatrix}
 \mathbb{I}_2 & \mathbb{G} \\
 \mathbb{G} & \mathbb{I}_2
 \end{pmatrix}, ~~~~
 \text{where} ~~~~
 \mathbb{G} \equiv
 \begin{pmatrix}
 0 & 0 \\
 g & 0
 \end{pmatrix}. 
\end{eqnarray}

{\bf Beam splitter}: We define the unitary for a beam splitter as
\begin{eqnarray}
\hat B_{jk}(\theta) = \exp \big[ -i \theta \big( \hat q_j \hat p_k  - \hat p_j \hat q_k \big) \big],  ~~~~~~ j \ne k, 
\end{eqnarray}
The symplectic matrix is 
\begin{eqnarray}\label{eq:BSSM}
B_{jk}(\theta) =
 \begin{pmatrix}
 \mathbb{I}_2 \cos \theta & -\mathbb{I}_2 \sin \theta \\
 \mathbb{I}_2 \sin \theta & ~~\, \mathbb{I}_2 \cos \theta
 \end{pmatrix}.
\end{eqnarray}

{\bf Displacement}: The displacement operator in the position quadrature is defined as
\begin{eqnarray}
 \hat X (s_j) = \exp (-i s_j \hat p_j)
\end{eqnarray}
and in the momentum quadrature it is defined as
\begin{eqnarray}
 \hat Z (s_j) = \exp (i s_j \hat q_j). 
\end{eqnarray}

The Wigner function is another quantity that can describe CV quantum states \cite{Leohnardt2010}. 
For Gaussian states, the Wigner functions are normalized Gaussian distributions and can be written in terms of the covariance matrix as
\begin{eqnarray}\label{eq:WignerFunction}
W (\boldsymbol \xi) = \frac{1}{(2 \pi)^M \sqrt{\text{det}\, \sigma}} \exp \bigg\{ - \frac{1}{2} {\boldsymbol \xi}^\top \sigma^{-1} {\boldsymbol \xi} \bigg\},
\end{eqnarray}
where $\boldsymbol{\xi} = (q_1, p_1, \cdots, q_M, p_M)^{\top}$ and $M$ is the number of modes. Under a unitary evolution $\hat U$, Eq. \eqref{eq:unitary}, the Wigner
function is transformed as 
\begin{eqnarray}\label{eq:WignerTransform}
W (\boldsymbol \xi) \rightarrow W^{\prime} (\boldsymbol \xi) = W \big[ S^{-1} (\boldsymbol \xi-\boldsymbol c) \big]. 
\end{eqnarray}

\subsection{Temporal CV Cluster State}

An ideal CV cluster state is a highly entangled multimode Gaussian state. It is generated by applying the controlled-Z gates, Eq. \eqref{eq:CZ}, to 
eigenstates of the momentum quadratures $\hat p_i$ \cite{Zhang2006}. The eigenstates of momentum quadratures are infinitely squeezed, thus requiring infinite energy. 
An approximate cluster state can be generated via replacing the momentum eigenstates by squeezed vacuum states in the momentum quadrature. 
Other methods of generating CV cluster states include linear-optics method \cite{Loock2007}, single-OPO method \cite{Menicucci2008, Flammia2009}, 
single-QND method \cite{Menicucci2010} and  a combination of these three \cite{Menicucci2011}. In this paper, we focus on the temporal CV cluster state \cite{Menicucci2011}
since it requires a small number of optical elements. A one-dimensional temporal CV cluster state with $10, 000$ entangled modes has been generated 
experimentally \cite{Yokoyama2013}, along with a one-million-mode version \cite{Yoshikawa2016}. A theoretical proposal for generating two-dimensional 
temporal CV cluster states \cite{Menicucci2011} is shown in Fig. \ref{fig:ECOpticalSetup}, with the error correction circuit (the shaded blue box) removed. 

\section{Single-mode error correction}\label{sec:SingleModeEC}

\subsection{General Formalism}

CV measurement-based quantum computing requires 
a resource state called a CV cluster state. In this section we consider implementing single-mode unitaries and it is known that a two-mode CV cluster state is adequate \cite{Alexander2014}.  An ideal two-mode CV cluster state can be created by applying 
a controlled-Z gate $\hat C_Z(t)$, Eq. (\ref{eq:CZ}), to two infinitely-squeezed momentum eigenstates. However, an ideal CV cluster state is unphysical
since it requires infinite energy. Here we generate an approximate two-mode cluster state by applying a controlled-Z gate to two single-mode squeezed vacuum states, 
both squeezed in the momentum quadrature, as shown in Fig. \ref{fig:SMU}. 
The interaction strength of the controlled-Z gate and the squeezing factor to the position quadrature of the squeezers are $t$ and $1/\sqrt{\epsilon}$, respectively. $t$ and
$\epsilon$ satisfy $t^2 + \epsilon^2 = 1$. Note that $\epsilon \rightarrow 0$ corresponds to infinite squeezing. An input mode couples with the cluster state via a beam splitter,
as shown in Fig. \ref{fig:SMU}, and a unitary can be implemented by performing homodyne measurements on two of the output modes. 
A specific unitary depends on the measurement angles $\theta_i$ ($i= 1, 3$), namely, the measurement quadratures $\hat b_{\theta_i}$ which are defined as
\begin{eqnarray}
\hat b_{\theta_i} = \hat p_i \cos \theta_i + \hat q_i \sin \theta_i. 
\end{eqnarray}
The quadratures $\hat q_1, \hat p_1, \hat q_3, \hat p_3$ are related to $\hat q_a, \hat p_a, \hat q_c, \hat p_c$ via a $50:50$ beam splitter, as shown in Fig. \ref{fig:SMU} 
(also see Appendix \ref{SMError} for details). It is shown that any single-mode unitary can be implemented by two measurement steps \cite{Alexander2014}. 
If the cluster state is ideal (infinite squeezing), the unitary 
can be implemented exactly. While for an approximate cluster state (finite squeezing) error
occurs and the state is distorted. This can be characterized by an error operator. 
Our main task is to correct the error by using the information of the input state. 

\subsection{Correcting Single-mode Errors}

We assume that the input state is a Gaussian state without a displacement. This is the case for some quantum algorithms, 
e.g., Gaussian Boson Sampling \cite{Hamilton2017, Bradler2017}, and the generalization to the case with a displacement is straightforward. 
The input state thus can be represented by a Wigner function 
$W_{\text{in}}(\boldsymbol \xi_a) = (2\pi \sqrt{{\rm det} \,\sigma_{{\rm in}}})^{-1} \exp ( - {\boldsymbol \xi_a}^\top \sigma^{-1}_{\text{in}} {\boldsymbol \xi_a}/2 )$,
where $\boldsymbol \xi_a = (q_a, p_a)^{\top}$ and $\sigma_{\text{in}}$ is the input covariance matrix. The unitary we want to implement is $\hat V_a$, so the target
covariance matrix is $\sigma_{\text{t}} = V_a \sigma_{\text{in}} V_a^{\top}$, where $V_a$ is the symplectic matrix of the mode $a$ denoted as 
$\hat V_a$. The target Wigner function is $W_{\text{target}}(\boldsymbol \xi_{c'}) = W_{\text{in}}(V_a^{-1} \boldsymbol \xi_{c'})$. 

\begin{figure}[ht!]
\includegraphics[width=8.3cm]{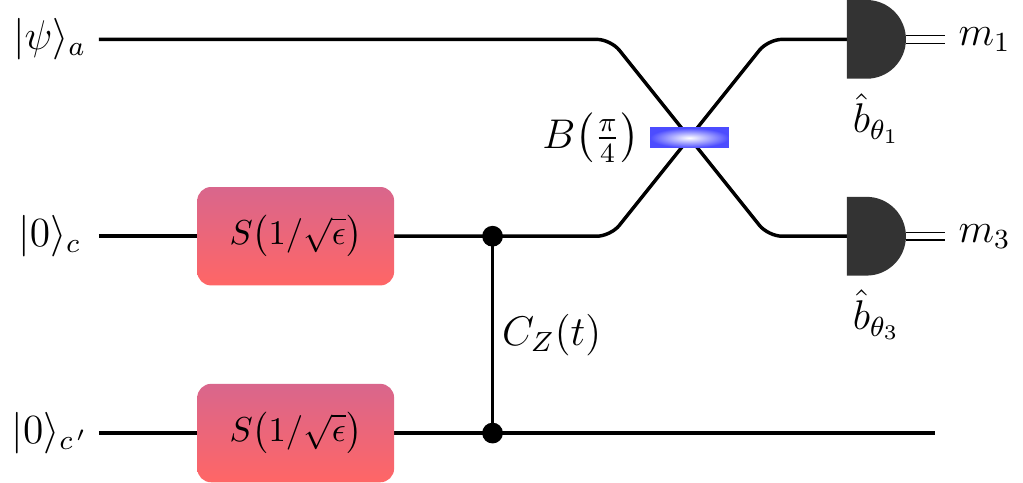}
\caption{ Measurement-based single-mode unitary implemented via a spatial two-mode CV cluster state. 
To implement a single-mode unitary, the input mode (mode $a$, with input state $|\psi \rangle_a$) 
is coupled to one (mode $c$) of the two modes of the CV cluster state
via a beam splitter $B(\pi/4)$. Two output modes are detected by homodyne detectors with measurement angles $\theta_1$ and $\theta_3$, respectively. 
After the homodyne measurements, one has to displace  the state in mode $c^{\,\prime}$ (not shown in the figure) 
according to the measurement outcomes $m_1$ and  $m_3$. } 
\label{fig:SMU}
\end{figure}

According to the protocol in Fig. \ref{fig:SMU}, the symplectic matrix $V_a$ is determined by the measurement angles via (see Appendix \ref{SMError} for details)
\begin{eqnarray}
V_a (\theta_1, \theta_3) = R_a \big(\theta_+/2  \big) S_a \big[ \tan (\theta_-/2)  \big] R_a \big(\theta_+/2  \big) R_a(\pi), \nonumber\\
\end{eqnarray}
where $\theta_{\pm} = \theta_1 \pm \theta_3$. Here $R_a(\theta)$ represents a phase shift with angle $\theta$ and $S_a(s)$ represents single-mode squeezing with
squeezing factor $s$. It is evident that various unitaries can be implemented by appropriately choosing the measurement angles. 
The displacements we need to apply after the homodyne detection are $\hat C^{\dag}(m_1, m_3) =  \hat Z^{\dag}(m_p) \hat X^{\dag}(m_q)$, where $\hat X$ and $\hat Z$ are
displacement operators in the $q$ and $p$ quadratures, respectively, and 
\begin{eqnarray}\label{eq:MoutcomeA}
m_q &=& \frac{\sqrt{2} \big( m_1 \sin \theta_3 + m_3 \sin \theta_1 \big)}{t \sin \theta_-}, \nonumber\\
m_p &=& - \frac{\sqrt{2} t \big( m_1 \cos \theta_3 + m_3 \cos \theta_1 \big)}{\sin \theta_-}. 
\end{eqnarray}
After the displacement correction, the output Wigner function $W_{\text{out}}$ is still different from the target Wigner function $W_{\text{target}}$ due to the effect of finite squeezing. 
We find
\begin{widetext}
\begin{eqnarray}\label{eq:Wigner-noAverage}
P(m_1, m_3) W_{\text{out}}({\boldsymbol \xi}_{c'})  
&=& 
\frac{2}{\pi^2}  \int \mathrm d x_q \int \mathrm d x_p \frac{1}{|\sin \theta_-|} W_{\text{in}} \big[V_a^{-1} ({\boldsymbol X} + {\boldsymbol x}) \big]  \nonumber\\
&&
\times \exp \bigg\{-{\boldsymbol x}^\top \Sigma_1 {\boldsymbol x} - ({\boldsymbol X}^\top + {\boldsymbol \gamma }^\top) \Sigma_2 {\boldsymbol x} 
- {\boldsymbol x}^\top \Sigma_2 ({\boldsymbol X} + {\boldsymbol \gamma})
- ({\boldsymbol X}^\top + {\boldsymbol \gamma}^\top) \Sigma_3 ({\boldsymbol X} + {\boldsymbol \gamma}) \bigg\}, 
\end{eqnarray}
\end{widetext}
where $P(m_1, m_3)$ is the probability with measurement outcomes $m_1$ and $m_3$, 
${\boldsymbol X} = S(t) {\boldsymbol \xi}_{c'}$, ${\boldsymbol \gamma} = S(t) \, (m_q, m_p)^{\top}$, 
${\boldsymbol x} = (x_q, x_p)^{\top}$, $\Sigma_1 = \frac{1}{\epsilon}\, \mathbb I_2$ and
\begin{eqnarray}\label{eq:Sigma}
 \Sigma_2 = 
 \begin{pmatrix}
 0 & 0 \\
 0 & \epsilon
 \end{pmatrix},
 ~~~~~~
 \Sigma_3 = 
 \begin{pmatrix}
 \epsilon/t^2 & 0 \\
 0 & \epsilon
 \end{pmatrix}. 
\end{eqnarray}
The measurement outcomes $m_1$ and $m_3$ are random numbers. Averaging over all measurement outcomes results in an averaged output state 
which contains noise due to the effect of finite squeezing \cite{Alexander2014}. For example, if the input state is a pure state, the averaged output state is generally a mixed state. 

If the input state covariance matrix ${\sigma_{\text{in}}}$ is known, we can straightforwardly perform the integration over $\boldsymbol x$ in Eq. \eqref{eq:Wigner-noAverage}. 
By performing an additional displacement, which depends not only on the measurement outcomes but also the input state and unitary (or the target output state), 
and integrating over all measurement outcomes, we obtain an output state with Wigner function
\begin{eqnarray}\label{eq:Wigner-output}
W({\boldsymbol \xi}_{c'})
&=&
\frac{1}{2\pi \sqrt{\text{det}\, \sigma}}  \exp \bigg\{ - \frac{1}{2}{\boldsymbol \xi}_{c'}^\top \sigma^{-1} \, {\boldsymbol \xi}_{c'} \bigg\},
\end{eqnarray}
where $\sigma$ is the actual output covariance matrix. The relation between the actual covariance matrix and the target covariance matrix is given by
\begin{eqnarray}\label{eq:relationTwoCM}
S^{-\top} (t) \, \sigma^{-1} S^{-1} (t) = \sigma_t^{-1} \big( \mathbb{I}_2 + \Delta_{\sigma} \big),
\end{eqnarray}
where $\Delta_{\sigma}$ characterizes the deviation of the actual covariance matrix to the target covariance matrix,
\begin{eqnarray}\label{eq:deviation}
\Delta_{\sigma} &=& 2 \sigma_t \Sigma_3
 - \big( \mathbb{I}_2 + 2 \sigma_t \Sigma_2 \big) \big( \mathbb{I}_2 + 2 \Sigma_1 \sigma_t \big)^{-1} \big( \mathbb{I}_2 +2 \sigma_t \Sigma_2 \big). \nonumber\\
\end{eqnarray}
The additional displacement to $\boldsymbol X$ is (see Appendix \ref{SMError} for details)
\begin{eqnarray}\label{FDiplacement}
{\boldsymbol D} = - \mathcal{A}^{-1}  \mathcal{B} {\boldsymbol \gamma}, 
\end{eqnarray}
where 
\begin{eqnarray}
\mathcal{A} &=& - \big(\sigma_t^{-1}+2 \Sigma_2 \big) \big(\sigma_t^{-1}+2 \Sigma_1 \big)^{-1} {\big(\sigma_t^{-1}+2 \Sigma_2 \big)} \nonumber\\
 && +\big(\sigma_t^{-1}+2 \Sigma_3 \big),   \nonumber\\
\mathcal{B} &=& 2 \Sigma_3 - \big(\sigma_t^{-1}+2 \Sigma_2 \big) \big(\sigma_t^{-1}+2 \Sigma_1 \big)^{-1} (2 \Sigma_2) . 
\end{eqnarray}
Note that in the infinite squeezing limit ($\epsilon \rightarrow 0, t \rightarrow 1$), $\Delta_{\sigma} \rightarrow 0$ and ${\boldsymbol D} \rightarrow 0$. 
This means in the infinite squeezing limit the actual output state is exactly the target state and no additional displacement is required. 
In order to recover the target covariance matrix, we need to apply another unitary, $\hat U_{\text{ec}}$, to the actual output state such that 
$\sigma_t = U_{\text{ec}} \, \sigma \, U_{\text{ec}}^{\top}$. We call the unitary $\hat U_{\text{ec}}$ the error correction unitary. 
 
 \subsection{Example 1}
 
To show how the error correction works, we now consider a concrete example. 
If the input state is a single-mode squeezed vacuum state with squeezing factor $\sqrt{s}$ and the unitary one wants to implement  is a $\pi/2$ phase shift, namely,
\begin{eqnarray}
\sigma_{\text{in}}
= \frac{1}{2}
\begin{pmatrix}
 s & 0 \\
 0 & 1/s
 \end{pmatrix},
 ~~~~~~
 V_a = 
 \begin{pmatrix}
 0 & -1 \\
 1 & \,\,\,\, 0
 \end{pmatrix}
 \end{eqnarray}
 then the target state is 
 \begin{eqnarray}\label{eq:TargetCM-SM}
 \sigma_t 
 =
 V_a \sigma_{\text{in}} V_a^\top
 = \frac{1}{2}
 \begin{pmatrix}
 1/s & 0 \\
 0 & s
 \end{pmatrix}.
\end{eqnarray}
By using Eq. \eqref{eq:deviation}, we find that the deviation is 
\begin{eqnarray}
\Delta_{\sigma} = 
\begin{pmatrix}
-\frac{\epsilon s^2 t^2 - \epsilon(1+\epsilon s)}{st^2(1+\epsilon s)} & 0 \\
 0 & \frac{\epsilon s^2 t^2 - \epsilon(1+\epsilon s)}{\epsilon + s}
 \end{pmatrix},
\end{eqnarray}
which is proportional to $\epsilon$ and vanishes when $\epsilon \rightarrow 0$. From Eq. \eqref{eq:relationTwoCM} the actual output state
covariance matrix can be calculated as 
\begin{eqnarray}
\sigma &=& 
 \frac{1}{2}
 \begin{pmatrix}
 \frac{ 1+\epsilon s}{\epsilon + s}  & 0 \\
 0 & \frac{\epsilon + s}{1+\epsilon s}
 \end{pmatrix}.
\end{eqnarray}
Evidently this is different from the target covariance matrix \eqref{eq:TargetCM-SM}. Note that $\sigma$ is a covariance matrix of a single-mode squeezed vacuum state. 
If $s<1$, the target state \eqref{eq:TargetCM-SM} is squeezed in the momentum quadrature with $s$ the minimal variance. 
The actual output state $\sigma$ is also squeezed in the momentum quadrature, however, the amount of squeezing decreases since $\frac{s +\epsilon}{1+\epsilon s} > s$.
Usually $\epsilon$ is assumed to be small (with a substantial amount of squeezing in the CV cluster state). In the limit of $s \rightarrow 0$, the minimal variance of the actual output 
state is approximately $s + \epsilon$,
which means after one measurement step the minimal variance increases $\epsilon$. To recover the target state, we have to further squeeze the actual output state. 
The symplectic matrix of the error correction unitary is found to be 
\begin{eqnarray}
 U_{\text{ec}}(\epsilon, s) = 
 \begin{pmatrix}
 \sqrt{ \frac{ 1+\epsilon/s}{1+\epsilon s}}  & 0 \\
 0 & \sqrt{\frac{1+\epsilon s}{1+\epsilon/s}}
 \end{pmatrix}. 
\end{eqnarray}

\section{Two-mode error correction}\label{sec:TwoModeEC}

\subsection{General Formalism}

\begin{widetext}

\begin{figure}[ht!]
\centering
\includegraphics[width=13cm]{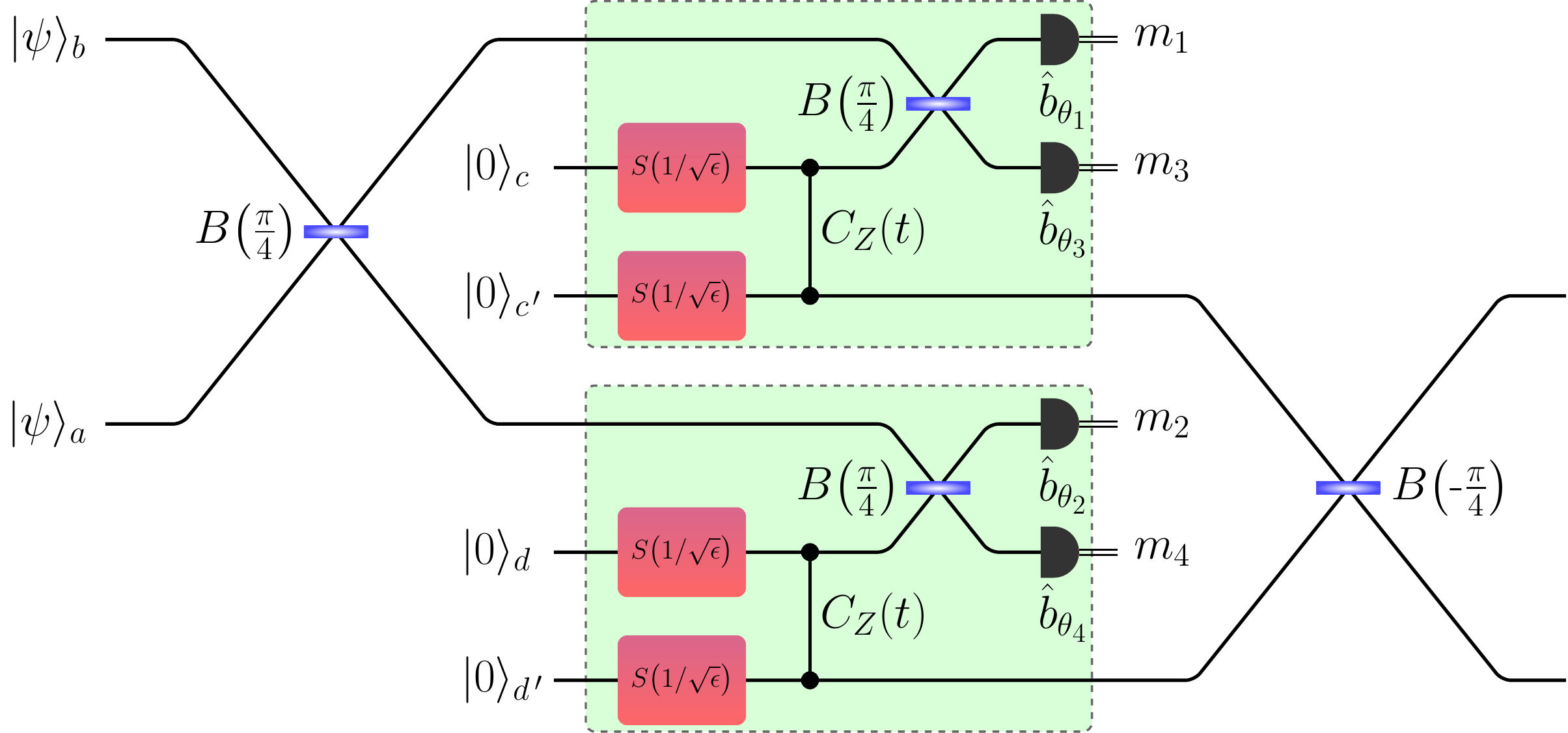}
\caption{ Measurement-based two-mode unitary implemented via a spatial four-mode CV cluster state. 
Two input modes $a$ and $b$ with input states $| \psi \rangle_a$ and $| \psi \rangle_b$ pass through
a beam splitter $B(\pi/4)$, the output states of which act as the input states of two measurement-based single-mode unitary circuits (two green shaded boxes). The top one 
is exactly the same as that shown in Fig. \ref{fig:SMU} and implements a unitary $\hat V_a(\theta_1, \theta_3)$; the bottom one is similar but with measurement quadratures 
$\hat b_{\theta_2}$ and $\hat b_{\theta_4}$, and implements a unitary $\hat V_b(\theta_2, \theta_4)$. 
The two output modes of these two measurement-based single-mode unitary circuits pass through another beam splitter
$B(-\pi/4)$, giving the final outputs. The implemented two-mode unitary is completely determined by the measurement angles $\theta_i$ ($i=1, \cdots, 4$). Note that
displacements depending on measurement outcomes $m_i$ have to be applied, which are not shown in the figure.  } 
\label{fig:TMU}
\end{figure}
\end{widetext}

A two-mode unitary can be implemented by using a quad-rail lattice cluster state \cite{Alexander2016}, which in fact is a universal CV cluster state that 
can implement arbitrary unitaries. In this section, we focus on correcting the errors induced by the finite squeezing in the quad-rail lattice cluster state. 
The two-mode unitary is implemented via four homodyne measurements,
the details of which were discussed in \cite{Alexander2016}. The implementation is shown to be equivalent to the scheme shown in Fig. \ref{fig:TMU}, where 
two measurement-based single-mode unitary schemes (see Fig. \ref{fig:SMU}) are sandwiched by two $50:50$ beam splitters: $B(\pi/4)$ and $B(-\pi/4)$.
Based on this observation, we can proceed with the help of the analysis in the single-mode case. 

\subsection{Correcting Two-mode Errors}

The single-mode unitary implemented by the circuit in the top green shaded box in Fig. \ref{fig:TMU} is $\hat V_a(\theta_1, \theta_3)$, while the bottom green shaded box implements
$\hat V_b(\theta_2, \theta_4)$, the symplectic matrix of which is given by
\begin{eqnarray}
V_b (\theta_2, \theta_4) &=&  R_b \big(\varphi_+/2  \big) S_b \big[ \tan (\varphi_-/2)  \big] R_b \big(\varphi_+/2  \big) R_b(\pi), \nonumber\\
\end{eqnarray}
where $\varphi_{\pm} \equiv \theta_2 \pm \theta_4$. Therefore the whole circuit in Fig. \ref{fig:TMU}  implements a two-mode unitary
\begin{eqnarray}
V
= B \bigg(-\frac{\pi}{4} \bigg) \bigg[ V_a(\theta_1, \theta_3) \oplus V_b (\theta_2, \theta_4)\bigg] B \bigg(\frac{\pi}{4} \bigg). 
\end{eqnarray}
Define ${\boldsymbol \xi}_2 = {\boldsymbol \xi}_{c'} \oplus {\boldsymbol \xi}_{d'}$, so the target Wigner function of the output state is
$W_{\text{target}}({\boldsymbol \xi}_{2}) = W_{\text{in}}( V^{-1}{\boldsymbol \xi}_{2})$. Similar to Eq. \eqref{eq:MoutcomeA}, we define
\begin{eqnarray}\label{eq:MoutcomeB}
\bar m_q &=& \frac{\sqrt{2} \big( m_2 \sin \theta_4 + m_4 \sin \theta_2 \big)}{t \sin \varphi_-}, \nonumber\\
\bar m_p &=& - \frac{\sqrt{2} t \big( m_2 \cos \theta_4 + m_4 \cos \theta_2 \big)}{\sin \varphi_-},
\end{eqnarray}
where $m_2$ and $m_4$ are the homodyne measurement outcomes, as shown in Fig. \ref{fig:TMU}. After the homodyne 
detection, we need to apply displacements $\hat C^{\dag}(m_1, m_3)$ and $\hat C^{\dag}(m_2, m_4)$ to the two modes right before the beam splitter $B(-\pi/4)$, respectively. 
Due to the finite squeezing in the cluster state, the Wigner function of the actual output state is different from $W_{\text{target}}({\boldsymbol \xi}_{2})$. 
We find that the actual Wigner function is given by (see Appendix \ref{TMEC} for details) 
\begin{widetext}
\begin{eqnarray}\label{eq:two-mode-Wigner}
P(m_1, m_2, m_3, m_4) W_{\text{out}}({\boldsymbol \xi}_2)  
&=& 
\frac{4}{\pi^4 |\sin \theta_- \, \sin \varphi_-|}  \int \mathrm d x_q \int \mathrm d x_p  \int \mathrm d y_q \int \mathrm d y_p 
\, W_{\text{in}} \bigg[ V^{-1} \big( \tilde{\boldsymbol X} + \tilde{\boldsymbol x}^{\prime} \big) \bigg] 
\nonumber\\
&&
\times \exp \bigg\{- \tilde {\boldsymbol x}^{\prime \top} \tilde \Sigma_1 \tilde {\boldsymbol x}^{\prime} - (\tilde {\boldsymbol X}^\top + \tilde {\boldsymbol \gamma}^\top) {\tilde \Sigma}_2^\top \tilde {\boldsymbol x}^{\prime} - \tilde {\boldsymbol x}^{\prime \top} \tilde \Sigma_2 (\tilde {\boldsymbol X} + \tilde {\boldsymbol \gamma})
- (\tilde {\boldsymbol X}^\top + \tilde {\boldsymbol \gamma}^\top) \tilde \Sigma_3 (\tilde {\boldsymbol X} + \tilde {\boldsymbol \gamma}) \bigg\}, \nonumber\\
\end{eqnarray}
\end{widetext}
where $P(m_1, m_2, m_3, m_4)$ is the probability of obtaining measurement outcomes $m_1, m_2, m_3, m_4$. 
Here $\tilde {\boldsymbol X} = \tilde S(t) {\boldsymbol \xi}_2$ with $\tilde S(t) = S(t) \oplus S(t)$, 
$\tilde{\boldsymbol x}^{\prime} = B(-\pi/4) \tilde {\boldsymbol x}$ with $\tilde {\boldsymbol x} = (x_q, x_p, y_q, y_p)^\top$, 
$\tilde {\boldsymbol \gamma} = \tilde S(t) B(-\pi/4) (m_q, m_p, \bar m_q, \bar m_p)^{\top}$, $\tilde \Sigma_1 = \frac{1}{\epsilon} \, \mathbb{I}_4$,
$\tilde \Sigma_2 = \Sigma_2 \oplus \Sigma_2$ and $\tilde \Sigma_3 = \Sigma_3 \oplus \Sigma_3$. 

Similar to the single-mode unitary case, if one averages over all measurement outcomes without knowing the information about the input state, one would end up with a 
state containing noise due to the effect of finite squeezing. For example, if the input state is pure, the output state is generally a mixed state. If we know the input state covariance
matrix $\sigma_{\text{in}}$, we can perform the integration over $\tilde{\boldsymbol x}$ in Eq. \eqref{eq:two-mode-Wigner}. 
We then apply an additional displacement to $\tilde {\boldsymbol X}$, which depends not only on the measurement outcomes but also the input state and unitary (or the target state), 
\begin{eqnarray}\label{FDiplacement-two}
\tilde {\boldsymbol D} = - \tilde{\mathcal{A}}^{-1}  \tilde{\mathcal{B}} \, \tilde {\boldsymbol \gamma},
\end{eqnarray}
where 
\begin{eqnarray}
\tilde{\mathcal{A}} &=&  - \big(\sigma_t^{-1}+2\tilde \Sigma_2 \big) \big(\sigma_t^{-1}+2\tilde \Sigma_1 \big)^{-1} {\big(\sigma_t^{-1}+2\tilde \Sigma_2 \big)}, \nonumber\\
 && + \big(\sigma_t^{-1}+2\tilde \Sigma_3 \big), \nonumber\\
\tilde{\mathcal{B}} &=& 2\tilde \Sigma_3 - \big(\sigma_t^{-1}+2\tilde \Sigma_2 \big) \big(\sigma_t^{-1}+2\tilde \Sigma_1 \big)^{-1} (2 \tilde \Sigma_2), 
\end{eqnarray}
and $\sigma_t = V \sigma_{\text{in}} V^{\top}$ is the target output covariance matrix. 
Finally, we average over all measurement outcomes and obtain an output state with Wigner function
\begin{eqnarray}\label{eq:Wigner-output}
W_{\text{out}}({\boldsymbol \xi}_2) 
= \frac{1}{(2\pi)^2 \sqrt{\text{det}\, \sigma_2}}  \exp \bigg\{ - \frac{1}{2}{\boldsymbol \xi}_2^{\prime \top} \sigma_2^{-1} \, {\boldsymbol \xi}_2 \bigg\},
\end{eqnarray}
where $\sigma_2$ is the actual output covariance matrix.  $\sigma_2$ is related to the target covariance matrix via
\begin{eqnarray}
\tilde S^{-\top} (t) \, \sigma_2^{-1} \tilde S^{-1} (t) = \sigma_t^{-1} \big( \mathbb{I}_4 + \tilde \Delta_{\sigma} \big),
\end{eqnarray}
where 
\begin{eqnarray}\label{Error:twomode-maintext}
\tilde \Delta_{\sigma} \equiv 2 \sigma_t \tilde \Sigma_3  - \big( \mathbb{I}_4 + 2\sigma_t \tilde \Sigma_2 \big) \big( \mathbb{I}_4 + 2 \tilde \Sigma_1 \sigma_t \big)^{-1} \big( \mathbb{I}_4 + 2 \sigma_t \tilde \Sigma_2 \big) \nonumber\\
\end{eqnarray}
characterizes the deviation from the target covariance matrix. Similarly, in the infinite squeezing limit ($\epsilon \rightarrow 0, t \rightarrow 1$), 
$\tilde \Delta_{\sigma} \rightarrow 0$ and $\tilde {\boldsymbol D} \rightarrow 0$. This means one can perfectly implement the unitary $\hat V$ without 
using the information of the input state and applying additional displacements. 

\subsection{Example 2}

\begin{figure}[ht!]
\includegraphics[width=8.3cm]{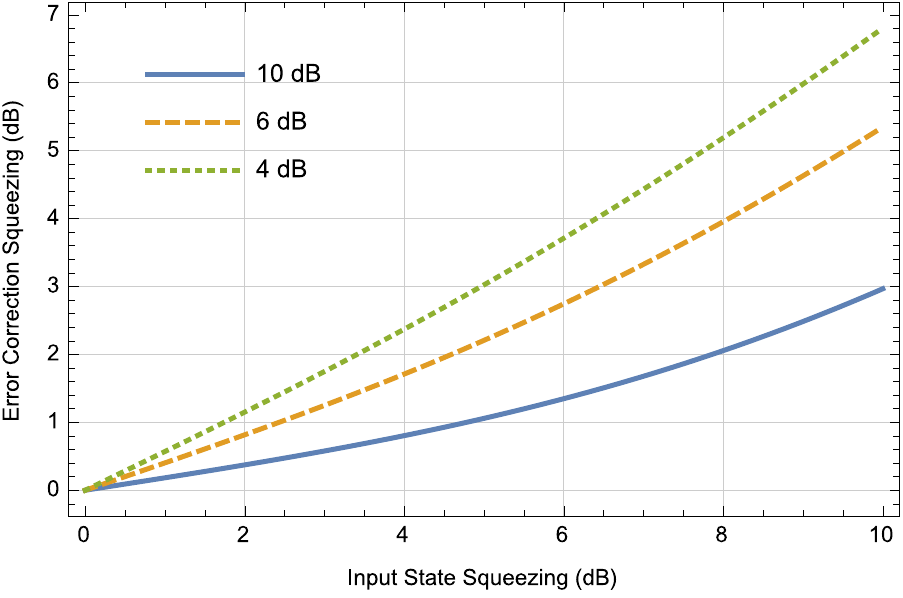}
\caption{ Amount of squeezing required in correcting two-mode errors versus the input state squeezing for a spatial CV cluster state with squeezing $4$ dB (green dotted line),
$6$ dB (orange dashed line) and $10$ dB (blue solid line). } 
\label{fig:EC-squeezing}
\end{figure}

To show the validity of the error correction scheme for two-mode unitaries, we consider a concrete example. 
Suppose that the two input states are two single-mode squeezed vacuum states which are squeezed in orthogonal directions and the implemented unitary
is a $50:50$ beam splitter. The target output state is a two-mode squeezed vacuum state, which we assume to be
\begin{eqnarray}
\sigma_t = \frac{1}{2}
 \begin{pmatrix}
 \cosh 2r \, \mathbb{I}_2 & \sinh 2r \, \mathbb{Z}_2 \\
 \sinh 2r \, \mathbb{Z}_2 & \cosh 2r \, \mathbb{I}_2
 \end{pmatrix},
\end{eqnarray}
where 
$
\mathbb{Z}_2 = 
\begin{pmatrix}
1 & 0 \\
0 & -1
\end{pmatrix}
$.
By substituting $\sigma_t$ into Eq. \eqref{Error:twomode-maintext} we find 
\begin{widetext}
\begin{eqnarray}
\mathbb{I}_4 + \tilde \Delta_{\sigma} &=& 
\begin{pmatrix}
 \frac{1+2\epsilon \cosh 2r + \epsilon^2 \cosh 4r}{t^2(1+\epsilon^2 + 2 \epsilon \cosh 2r)} & 0 &  \frac{2\epsilon \sinh 2r (1+\epsilon \cosh 2r)}{t^2(1+\epsilon^2 + 2 \epsilon \cosh 2r)} & 0 \\
 0 &  \frac{t^2(1+2\epsilon \cosh 2r + \epsilon^2 \cosh 4r)}{1+\epsilon^2 + 2 \epsilon \cosh 2r} & 0 & -\frac{2t^2 \epsilon \sinh 2r (1+\epsilon \cosh 2r)}{1+\epsilon^2 + 2 \epsilon \cosh 2r} \\
 \frac{2\epsilon \sinh 2r (1+\epsilon \cosh 2r)}{t^2(1+\epsilon^2 + 2 \epsilon \cosh 2r)} & 0&  \frac{1+2\epsilon \cosh 2r + \epsilon^2 \cosh 4r}{t^2(1+\epsilon^2 + 2 \epsilon \cosh 2r)} & 0 \\
 0 & -\frac{2t^2 \epsilon \sinh 2r (1+\epsilon \cosh 2r)}{1+\epsilon^2 + 2 \epsilon \cosh 2r} & 0&  \frac{t^2(1+2\epsilon \cosh 2r + \epsilon^2 \cosh 4r)}{1+\epsilon^2 + 2 \epsilon \cosh 2r} 
 \end{pmatrix} \nonumber\\
 \nonumber\\
 &=&
 \tilde S^{-1} (t) 
 \begin{pmatrix}
 \mathcal{J} \, \mathbb{I}_2 & \mathcal{K} \, \mathbb{Z}_2 \\
 \mathcal{K} \, \mathbb{Z}_2 & \mathcal{J} \, \mathbb{I}_2
 \end{pmatrix}
 \tilde S^{-\top} (t),
\end{eqnarray}
\end{widetext}
where we have defined
\begin{eqnarray}
\mathcal{J} &=&  \frac{1+2\epsilon \cosh 2r + \epsilon^2 \cosh 4r}{1+\epsilon^2 + 2 \epsilon \cosh 2r}, \nonumber\\
\mathcal{K} &=& \frac{2\epsilon \sinh 2r (1+\epsilon \cosh 2r)}{1+\epsilon^2 + 2 \epsilon \cosh 2r}.
\end{eqnarray} 
It can be shown straightforwardly that $\tilde S(t) \sigma_t = \sigma_t \tilde S(t)$, so the target covariance matrix $\sigma_t$ can be recovered from the actual 
output covariance matrix $\sigma_2$ via
\begin{eqnarray}
\sigma_t = 
\begin{pmatrix}
 \mathcal{J} \, \mathbb{I}_2 & \mathcal{K} \, \mathbb{Z}_2 \\
 \mathcal{K} \, \mathbb{Z}_2 & \mathcal{J} \, \mathbb{I}_2
 \end{pmatrix}
 \sigma_2.
\end{eqnarray}
Note that $\mathcal{J}^2 - \mathcal{K}^2 = 1$, so we can define 
\begin{eqnarray}
\mathcal{J} &=&  \cosh 2\alpha, ~~~~~~
\mathcal{K} = \sinh 2\alpha,
\end{eqnarray}
such that
\begin{eqnarray}
\sigma_t &=& 
\begin{pmatrix}
 \cosh \alpha \, \mathbb{I}_2 &  \sinh \alpha \, \mathbb{Z}_2 \\
  \sinh \alpha \, \mathbb{Z}_2 &  \cosh \alpha \, \mathbb{I}_2
 \end{pmatrix}
 \sigma_2 
 \begin{pmatrix}
 \cosh \alpha \, \mathbb{I}_2 &  \sinh \alpha \, \mathbb{Z}_2 \\
  \sinh \alpha \, \mathbb{Z}_2 &  \cosh \alpha \, \mathbb{I}_2
 \end{pmatrix}. \nonumber\\
\end{eqnarray}
Therefore the target covariance matrix can be recovered by applying another two-mode squeezer with squeezing parameter $\alpha$ to the actual output state. 
Fig. \ref{fig:EC-squeezing} shows the relation between the error correction squeezing parameter $\alpha$ and the input state squeezing parameter $r$ for some 
particular numbers of squeezing in the cluster state. Note that the squeezing has been converted into the units of dB using $10 \log_{10} (e^{2r}) \approx 8.686 \,r$ 
where $r$ is the squeezing parameter. 

\section{Multimode error correction}\label{sec:MultiModeEC}

In CV quantum computation, a typical circuit usually contains a large number of modes, implementing multimode unitaries. 
An arbitrary multimode unitary can always be decomposed into a collection of single-mode and two-mode unitaries. The error correction scheme for single-mode and 
two-mode unitaries discussed in Secs. \ref{sec:SingleModeEC} and \ref{sec:TwoModeEC} thus can be easily fitted into a multimode circuit. 
We showed that there exists no resource advantage if the multimode unitaries are implemented by spatial CV cluster states.  
However, as discussed below, in the time domain cluster states practical advantages can be obtained as the resources stay fixed for arbitrary amounts of input modes.

\subsection{Error Correction Optical Setup}


\begin{figure*}[ht!]
\centering
\includegraphics[width=18cm]{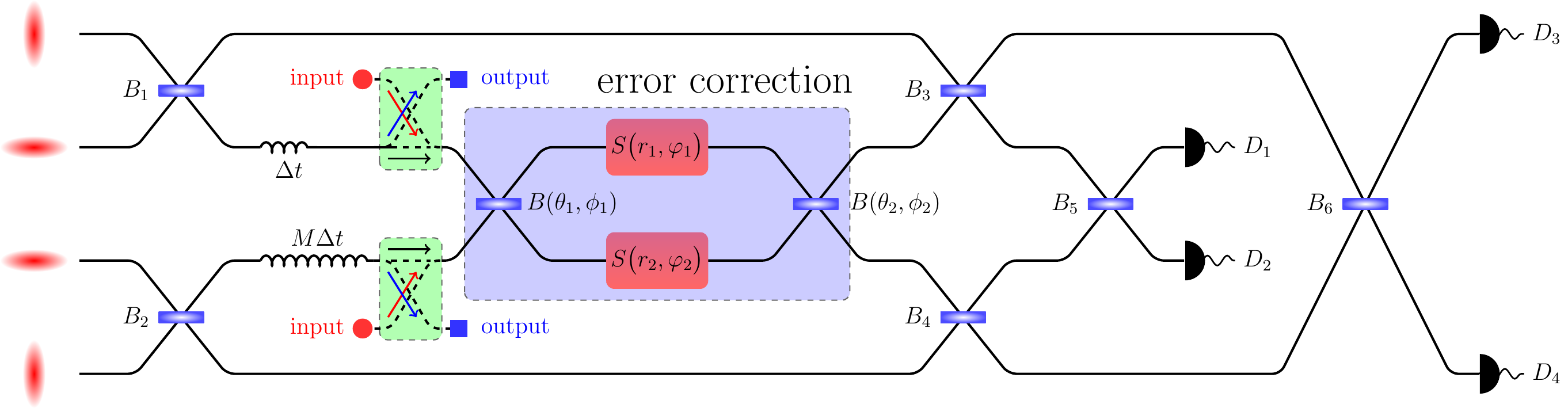}
\caption{ The error correction circuit (blue shaded box) for multimode Gaussian unitaries implemented via a temporal CV cluster state 
consists of two beam splitters, $B(\theta_1, \phi_1)$ and $B(\theta_2, \phi_2)$, and two single-mode
squeezers, $S(r_1, \varphi_1)$ and $S(r_2,  \varphi_2)$ (note that we have absorbed the phase shifts into the beam splitters and squeezers to simplify the circuit). 
The parameters of these beam splitters and squeezers are adjusted to corrects errors imposed on implemented two-mode unitaries. 
The remaining part of the optical setup generates two-dimensional CV cluster states  \cite{Menicucci2011} and implements universal quantum computation 
with addition of non-Gaussian gates \cite{Alexander2017UQC}, e.g., cubic-phase gates. $B_i \,(i = 1, \cdots,6)$ are six $50:50$ beam splitters, 
$D_i \, (i = 1, 2, 3, 4)$ are four homodyne detectors. Four single-mode squeezed pulses (left most) are generated continuously with time interval $\Delta t$ and are injected 
into the circuit. A pair of two-mode squeezed states are produced after the beam splitter $B_1$ and $B_2$. One pulse of the top two-mode squeezed state is delayed by 
$\Delta t$, while one pulse of the bottom two-mode squeezed state is delayed by $M \Delta t$, where $M$ is an integer and can thought of as the depth of the circuit. 
If the error correction circuit is absent and no homodyne detection is performed, a temporal two-dimensional cluster state is generated after the beam splitters 
$B_3$, $B_4$, $B_5$ and $B_6$ \cite{Menicucci2011}. To do quantum computation, two switches (green boxes) have to be introduced to couple the input states into the cluster 
state or to readout the output states after the computation \cite{Alexander2017MBLO}. 
At the beginning, the switches are turned on to let the input states couple into the cluster state and then turned off. Homodyne measurements are performed to implement a 
two-mode unitary each time, depending on the measurement quadratures of the homodyne detection. Error correction follows the homodyne measurements to 
correct the error induced by the finite squeezing in the input squeezed pulses (note that displacements are also required, which are not shown in the figure). After all computations
are done, the switches are turned on such that one can readout or detect the output states. 
 } 
\label{fig:ECOpticalSetup}
\end{figure*}

From the single-mode and two-mode cases, we can summarize the procedure of the error correction scheme as follows. Before applying the unitary, we have to know the input
state, e.g., the input Wigner function. This is motivated from the observation that for most quantum algorithms the input states are known. 
Since we also know the unitary that is going to be implemented, we can straightforwardly calculate the expected target output Wigner function.
This tells us the information about the additional displacement and the error that we have to correct. Appropriate optical elements (a single-mode squeezer and several 
phase shifters for single-mode unitary error correction; two single-mode squeezers, two beam splitters and several phase shifters for two-mode unitary error correction)
are introduced to build an error correction circuit. We then perform homodyne measurements, displace the output state (standard and additional displacements) and correct the
actual output state by passing it through the error correction circuit. Finally, the target state is obtained, namely, the unitary has been perfectly implemented. 

If one is only concerned with single-mode and two-mode unitaries, the proposed error correction scheme is not necessary because it costs too much in terms of resources. 
For example, one can directly prepare an almost perfect beam splitter instead of implementing a virtual beam splitter via measuring a  cluster state and then correcting the error
using another two beam splitters and two single-mode squeezers. Obviously, there is no point in doing that. We can generalize the error correction scheme to multimode
unitaries which can always be decomposed into a collection of single-mode and two-mode unitaries. If the multimode unitaries are implemented by spatial CV cluster states,
we have to add an error correction circuit after each single-mode or two-mode unitary. The number of required optical elements is thus huge, indicating that there exists no advantage 
to using this error correction scheme for the spatial CV clusters. 

However, this error correction scheme shows advantages when one considers a multimode circuit implemented by a temporal CV cluster 
state \cite{Yokoyama2013}. 
It was proposed that a measurement-based linear optical circuit can be 
implemented using a temporal CV cluster state \cite{Alexander2017MBLO}. The advantage of the measurement-based protocol is that a circuit with a large number of modes 
can be achieved with a small number of optical elements. For example, a one-million-mode CV cluster state has been generated  \cite{Yoshikawa2016}.
It is promising to generalize the measurement-based protocol to generate two-dimensional cluster states such that
multimode unitaries can be implemented. With a small number of additional optical elements, the proposed error correction scheme can be well fitted into the measurement-based
multimode circuit. 

Fig. \ref{fig:ECOpticalSetup} shows a time domain optical setup that generates two-dimensional cluster states and implements 
multimode unitaries \cite{Alexander2017MBLO} with the error correction circuit included (the blue shaded box). An arbitrary multimode unitary can be decomposed into 
a set of single-mode and two-mode unitaries. The circuit shown in Fig.~\ref{fig:ECOpticalSetup} implements either a two-mode unitary or two single-mode unitaries by 
performing four homodyne measurements at once. Therefore we only need to correct a two-mode unitary or two single-mode unitaries each time. The optical elements 
required to perform these error corrections are at most two single-mode squeezers, two beam splitters and several phase shifters. By continuously generating four single-mode
squeezed states and performing homodyne measurements, an error corrected multimode unitary is implemented. Note that at each step the unitary and the input
state are different, so the parameters (the amount of squeezing, transmission coefficient, phase etc.) of the optical elements in the error correction circuit should be adjustable.  

\subsection{Classical Resources}

To achieve error correction for a multimode circuit, one has to keep track of the Wigner function through the whole computational process. In particular, one has to
store the information of the Wigner function before and after each unitary, and calculate the evolution of the Wigner function based on this unitary. 
Consider an $M$-mode circuit, the size of the covariance matrix is $2M \times 2M$, which means one needs to store at most $(2M)^2$ complex numbers in a classical computer. 
The computation proceeds step by step with each step implementing a two-mode unitary. The submatrix of the covariance matrix corresponding to these two modes
is changed, as well as the correlations between these two modes and other modes. This means $4(2M-1)$ elements of the covariance matrix are changed after a two-mode
unitary. From this conservative estimate, the size of the required memory in a classical computer grows polynomially in the size of the circuit. This means that the Wigner function
can be tracked efficiently.

\section{Conclusion}\label{sec:Conclusion}

We proposed an error correction scheme for measurement-based quantum computation based on the temporal continuous-variable cluster states. To perform error correction due 
to the effect of finite squeezing in the cluster states, we used the information of the input states, which is typical for quantum computing algorithms.
Furthermore, the Wigner function has to be kept tracked of during the whole computational process. 
The parameters of the optical elements in the error correction circuit are adjusted according to the input state and unitary. For a computation based on temporal cluster states,
we find that a small number of optical elements is adequate: two beam splitters, two single-mode squeezers and several phase shifters. 
We note that future work would include analyzing the effects of other types of errors, such as those from experimental imperfections. 

The challenge of this error correction
scheme is the online squeezing, which is more difficult to achieve experimentally than the offline squeezing. 
However, a certain amount of online squeezing has been achieved experimentally \cite{Miwa2014, Marshall2016}
and we show than the typical amount of online squeezing is not so high if the squeezing in the cluster state is above a certain level (see Fig. \ref{fig:EC-squeezing}). 
Therefore the proposed error correction scheme is promising if both the amount of online squeezing and the squeezing in the cluster states can be improved in the near future. 

Finally, to achieve universal quantum computation, non-Gaussian states or non-Gaussian gates, e.g., cubic-phase gates, are required. 
Therefore, the error correction involving non-Gaussian states or non-Gaussian unitaries is paramount. It is important to explore whether the proposed error correction scheme 
can be generalized to the non-Gaussian regime. We leave this for future work. 

\section*{Acknowledgements}

We thank Patrick Rebentrost for useful discussions.

\appendix
\vspace{0.5cm}

\begin{widetext}

\section{Single-mode Errors Due to Finite Squeezing}\label{SMError}

%

In this appendix, we focus on single-mode errors due to the finite squeezing in the CV cluster states. 
We first derive the output Wigner function $W_{\rm out}$, Eq. \eqref{eq:Wigner-noAverage}, of the measurement-based unitary circuit shown in Fig. \ref{fig:SMU},
then discuss in detail how to correct the single-mode errors. 

\subsection{Wigner Function Before Homodyne Measurement}

We follow the evolution of the Wigner function from left to right in the circuit shown in Fig. \ref{fig:SMU}. 
Assume that the input state is a Gaussian state and is represented by a Wigner function $W_{\text{in}}(q_a, p_a)$. 
The Wigner function for two independent single-mode squeezed states with squeezing factor $1/\sqrt{\epsilon}$, which are both squeezed in the momentum quadrature, is 
\begin{eqnarray}
W_{\text{SS}}(q_c, p_c, q_{c'}, p_{c'}) = G_{1/\epsilon}(q_c)  G_{\epsilon}(p_c) G_{1/\epsilon}(q_{c'}) G_{\epsilon}(p_{c'}),
\end{eqnarray}
where $G_y(x)$ is a Gaussian function,
\begin{eqnarray}\label{eq:Gaussian}
G_y(x) = \frac{1}{\sqrt{\pi y}} \exp \bigg\{-\frac{x^2}{y} \bigg\}. 
\end{eqnarray}
By using Eqs. \eqref{eq:CZSM} and \eqref{eq:WignerTransform}, we can obtain the Wigner function of an approximate CV cluster state, 
\begin{eqnarray}
W_{\text{CS}}(q_c, p_c, q_{c'}, p_{c'}) = G_{1/\epsilon}(q_c)  G_{\epsilon}(p_c-t  q_{c'}) G_{1/\epsilon}(q_{c'}) G_{\epsilon}(p_{c'} - t q_c)
\end{eqnarray}
The overall Wigner function before the beam splitter is 
\begin{eqnarray}
W_{\text{in}}(q_a, p_a) W_{\text{CS}}(q_c, p_c, q_{c'}, p_{c'}) = W_{\text{in}}(q_a, p_a) G_{1/\epsilon}(q_c)  G_{\epsilon}(p_c-t  q_{c'}) G_{1/\epsilon}(q_{c'}) G_{\epsilon}(p_{c'} - t q_c).
\end{eqnarray}
We denote the quadratures after the beam splitter as $( \hat q_1, \hat p_1, \hat q_3, \hat p_3 )$, and they are related to the quadratures before the beam splitter 
$( \hat q_a, \hat p_a, \hat q_c, \hat p_c )$ via Eq. \eqref{eq:BSSM} with $\theta = \pi/4$, namely,
\begin{eqnarray}
 \begin{pmatrix}
  \hat q_3 \\
  \hat q_1
 \end{pmatrix}
 = \frac{1}{\sqrt{2}}
 \begin{pmatrix}
  1 & - 1 \\
  1 &  \,\,\,\,\,1 
 \end{pmatrix}
  \begin{pmatrix}
  \hat q_a \\
  \hat q_c
 \end{pmatrix}, 
 ~~~~~~
  \begin{pmatrix}
  \hat p_3 \\
  \hat p_1
 \end{pmatrix}
 = \frac{1}{\sqrt{2}}
 \begin{pmatrix}
  1 & - 1 \\
  1 &  \,\,\,\,\,1 
 \end{pmatrix}
  \begin{pmatrix}
  \hat p_a \\
  \hat p_c
 \end{pmatrix}.
\end{eqnarray}
Therefore after the beam splitter (before performing the homodyne measurement), the overall Wigner function becomes
\begin{eqnarray}\label{eq:Wigner-noHM}
W_{\text{in}}\bigg[ \frac{1}{\sqrt{2}} ( q_1 + q_3 ), \frac{1}{\sqrt{2}}( p_1 + p_3) \bigg] G_{1/\epsilon}\bigg[ \frac{1}{\sqrt{2}}( q_1 - q_3)\bigg]  
G_{\epsilon}\bigg[ \frac{1}{\sqrt{2}}( p_1 - p_3)-t  q_{c'} \bigg] G_{1/\epsilon}(q_{c'}) G_{\epsilon} \bigg[ p_{c'} - \frac{t}{\sqrt{2}}( q_1 - q_3)\bigg].
\end{eqnarray}

\subsection{Homodyne Measurement}

Homodyne detector measures the quadrature of the optical field, e.g., the position quadrature $\hat q$ or momentum quadrature $\hat p$. 
More generally, a phase shift can be inserted before the detector such that an arbitrary quadrature is measured. 
Suppose that $\hat b_{\theta}$ is the quadrature we want to measure and $\hat c_{\theta}$ is its conjugate quadrature,
$[\hat c_{\theta}, \hat b_{\theta}] = i$, then
\begin{eqnarray}
 \begin{pmatrix}
  \hat c_{\theta} \\
  \hat b_{\theta}
 \end{pmatrix}
 =
 \begin{pmatrix}
  \cos \theta & - \sin \theta \\
   \sin \theta &  \,\,\,\,\, \cos \theta 
 \end{pmatrix}
 \begin{pmatrix}
  \hat q \\
  \hat p
 \end{pmatrix}. 
\end{eqnarray} 
In the circuit shown in Fig. \ref{fig:SMU}, we measure the quadratures $\hat b_{\theta_1}$ and $\hat b_{\theta_3}$, respectively. If we measure $\hat b_{\theta_1}$
and obtain an outcome $m_1$,  we get a constraint: $p_1 \cos \theta_1 + q_1 \sin \theta_1 = m_1$. The possible measurement outcome (we do not measure it actually) of its conjugate 
quadrature $\hat c_{\theta_1}$ is independent and could be any real values, which is denoted as $\tau_1$ and satisfies $-p_1 \sin \theta_1 + q_1 \cos \theta_1 = \tau_1$. 
We have 
\begin{eqnarray}\label{eq:constraint-1}
 \begin{pmatrix}
  \tau_1 \\
  m_1
 \end{pmatrix}
 =
 \begin{pmatrix}
  \cos \theta_1 & - \sin \theta_1 \\
   \sin \theta_1 &  \,\,\,\,\, \cos \theta_1
 \end{pmatrix}
 \begin{pmatrix}
  q_1 \\
  p_1
 \end{pmatrix}. 
\end{eqnarray}
Similarly, for the other mode, 
\begin{eqnarray}\label{eq:constraint-3}
 \begin{pmatrix}
  \tau_3 \\
  m_3
 \end{pmatrix}
 =
 \begin{pmatrix}
  \cos \theta_3 & - \sin \theta_3 \\
   \sin \theta_3 &  \,\,\,\,\, \cos \theta_3
 \end{pmatrix}
 \begin{pmatrix}
  q_3 \\
  p_3
 \end{pmatrix}. 
\end{eqnarray}
where $m_3$ is the measurement outcome of $\hat b_{\theta_3}$ and  $\tau_3$ is the possible measurement outcome of $\hat c_{\theta_3}$. 
Substituting the constraints \eqref{eq:constraint-1} and \eqref{eq:constraint-3} into the overall Wigner function \eqref{eq:Wigner-noHM} and integrating over 
$\tau_1$ and $\tau_3$, we find
\begin{eqnarray}\label{eq:Wigner-2}
P(m_1, m_3) W_{\text{out}}(q_{c'}, p_{c'}) 
&=& \int \mathrm d \tau_1 \int \mathrm d \tau_3 ~W_{\text{in}}\bigg[ \frac{1}{\sqrt{2}} \big( m_1 \sin \theta_1+ m_3 \sin \theta_3 \big) + \frac{1}{\sqrt{2}} \big( \tau_1 \cos \theta_1+ \tau_3 \cos\theta_3 \big),  \nonumber\\
&&\frac{1}{\sqrt{2}} \big( m_1 \cos \theta_1+ m_3 \cos \theta_3 \big) - \frac{1}{\sqrt{2}} \big( \tau_1 \sin \theta_1+ \tau_3 \sin \theta_3 \big) \bigg] \nonumber\\
&& \times G_{1/\epsilon}\bigg[ \frac{1}{\sqrt{2}} \big( m_1 \sin \theta_1- m_3 \sin \theta_3 \big) + \frac{1}{\sqrt{2}} \big( \tau_1 \cos \theta_1- \tau_3 \cos\theta_3 \big) \bigg]
G_{1/\epsilon}(q_{c'}) \nonumber\\
&&\times G_{\epsilon}\bigg[ \frac{1}{\sqrt{2}} \big( m_1 \cos \theta_1- m_3 \cos \theta_3 \big) - \frac{1}{\sqrt{2}} \big( \tau_1 \sin \theta_1- \tau_3 \sin \theta_3 \big)-t  q_{c'} \bigg]
\nonumber\\
&&\times G_{\epsilon} \bigg[ p_{c'} - \frac{t}{\sqrt{2}}\big( m_1 \sin \theta_1- m_3 \sin \theta_3 \big) - \frac{t}{\sqrt{2}} \big( \tau_1 \cos \theta_1- \tau_3 \cos\theta_3 \big) \bigg], 
\end{eqnarray}
where $P(m_1, m_3)$ is the probability of registering measurement outcomes $m_1$ and $m_3$ simultaneously. In the case of $\theta_1 \neq \theta_3$, 
by changing the integration variables and performing some algebraic calculations we find
\begin{eqnarray}\label{eq:Wigner-4}
&&P(m_1, m_3) W_{\text{out}}(q_{c'}, p_{c'})  \nonumber\\
&=& 
\int \mathrm d x_q \int \mathrm d x_p \frac{2}{|\sin \theta_-|} W_{\text{in}}\bigg[ -( x_q +  t  q_{c'}) \bigg( \frac{2 \cos \theta_1 \cos \theta_3}{\sin \theta_-} \bigg) - \bigg( x_p + \frac{p_{c'}}{t} \bigg) \bigg( \frac{\sin \theta_+}{\sin \theta_-} \bigg) 
+ \frac{\sqrt{2} \big( m_1 \cos \theta_3- m_3 \cos \theta_1 \big) }{\sin \theta_-},  \nonumber\\
&&
+ \bigg( x_q + t  q_{c'} \bigg) \bigg( \frac{\sin \theta_+}{\sin \theta_-} \bigg) + 
\bigg( x_p + \frac{p_{c'}}{t} \bigg) \bigg( \frac{2 \sin \theta_1 \sin \theta_3}{\sin \theta_-} \bigg) 
 - \frac{\sqrt{2} \big( m_1 \sin \theta_3- m_3 \sin \theta_1 \big) }{\sin \theta_-} \bigg] \nonumber\\
&& 
\times G_{1/\epsilon}\bigg( x_p + \frac{p_{c'}}{t} \bigg)
G_{1/\epsilon}(q_{c'}) G_{\epsilon}( x_q )
G_{\epsilon} ( - t x_p).
\end{eqnarray}
We further define $m_q$ and $m_p$ as
\begin{eqnarray}
m_q &=& \frac{\sqrt{2} \big( m_1 \sin \theta_3 + m_3 \sin \theta_1 \big)}{t \sin \theta_-}, \nonumber\\
m_p &=& - \frac{\sqrt{2} t \big( m_1 \cos \theta_3 + m_3 \cos \theta_1 \big)}{\sin \theta_-}
\end{eqnarray}
so that the Wigner function can be rewritten as
\begin{eqnarray}\label{eq:Wigner-6}
&&P(m_1, m_3) W_{\text{out}}(q_{c'}, p_{c'})  \nonumber\\
&=& 
\int \mathrm d x_q \int \mathrm d x_p \frac{2}{|\sin \theta_-|} W_{\text{in}}\bigg[ -\bigg( x_q +  t  \big(q_{c'} - m_q \big) \bigg) \bigg( \frac{ 2 \cos \theta_1 \cos \theta_3}{\sin \theta_-} \bigg) - \bigg( x_p + \frac{\big( p_{c'} - m_p \big) }{t} \bigg) \bigg( \frac{\sin \theta_+}{\sin \theta_-} \bigg),  \nonumber\\
&&
\bigg( x_q + t \big(q_{c'} - m_q \big) \bigg) \bigg( \frac{\sin \theta_+}{\sin \theta_-} \bigg) + 
\bigg( x_p + \frac{\big( p_{c'} - m_p \big)}{t} \bigg) \bigg( \frac{2 \sin \theta_1 \sin \theta_3}{\sin \theta_-} \bigg) \bigg] \nonumber\\
&& 
\times G_{1/\epsilon}\bigg( x_p + \frac{p_{c'}}{t} \bigg)
G_{1/\epsilon}(q_{c'}) G_{\epsilon}\big(x_q \big)
G_{\epsilon} \big( - t x_p \big). 
\end{eqnarray}
The two arguments of the Wigner function $W_{\text{in}}$ can be considered as a two-component vector, which we find can be written as
\begin{eqnarray}
-\frac{1}{\sin \theta_-} 
 \begin{pmatrix}
 2 \cos \theta_1 \cos \theta_3 & \sin \theta_+\\
 -\sin \theta_+& - 2 \sin \theta_1 \sin \theta_3
 \end{pmatrix}
 \bigg\{
 \begin{pmatrix}
 t & 0 \\
 0 & 1/t
 \end{pmatrix}
 \bigg[
 \begin{pmatrix}
 q_{c'} \\
 p_{c'} 
 \end{pmatrix}
 -
 \begin{pmatrix}
 m_q \\
 m_p
 \end{pmatrix}
 \bigg]
 +
 \begin{pmatrix}
 x_q \\
 x_p
 \end{pmatrix}
 \bigg\}.
\end{eqnarray}
We define 
\begin{eqnarray}
V_a^{-1} (\theta_1, \theta_3) = 
-\frac{1}{\sin \theta_-} 
 \begin{pmatrix}
 2 \cos \theta_1 \cos \theta_3 & \sin \theta_+\\
 -\sin \theta_+& - 2 \sin \theta_1 \sin \theta_3
 \end{pmatrix}
 =
- \frac{1}{\sin \theta_-} 
 \begin{pmatrix}
 \cos \theta_+ + \cos \theta_- & \sin \theta_+\\
 -\sin \theta_+&  \cos \theta_+ - \cos \theta_-
 \end{pmatrix}
\end{eqnarray}
and find that it can be decomposed into
\begin{eqnarray}
V_a^{-1} (\theta_1, \theta_3) &=& -
 \begin{pmatrix}
 \cos \big(\frac{\theta_+}{2} \big) & \sin \big(\frac{\theta_+}{2} \big) \\
 \\
 -\sin \big(\frac{\theta_+}{2} \big) & \cos \big(\frac{\theta_+}{2} \big)
 \end{pmatrix}
  \begin{pmatrix}
 \tan^{-1} \big(\frac{\theta_-}{2} \big) & 0 \\
 \\
 0 & \tan \big(\frac{\theta_-}{2} \big)
 \end{pmatrix}
  \begin{pmatrix}
 \cos \big(\frac{\theta_+}{2} \big) & \sin \big(\frac{\theta_+}{2} \big) \\
 \\
 -\sin \big(\frac{\theta_+}{2} \big) & \cos \big(\frac{\theta_+}{2} \big)
 \end{pmatrix} \nonumber \\
 \nonumber\\
 &=&
 R^{-1}(\pi)R^{-1} \big(\theta_+/2  \big) S^{-1} \big[ \tan (\theta_-/2)  \big] R^{-1} \big(\theta_+/2  \big). 
\end{eqnarray}
This shows that choosing homodyne measurement angles $\theta_1$ and $\theta_3$ implements a unitary $\hat V (\theta_+, \theta_-)$, the symplectic matrix of which is
\begin{eqnarray}
V_a (\theta_1, \theta_3) = R \big(\theta_+/2  \big) S \big[ \tan (\theta_-/2)  \big] R \big(\theta_+/2  \big) R(\pi). 
\end{eqnarray}
In addition, the state is displaced by $\hat X(m_q)$ and $\hat Z(m_p)$. More importantly, the state is distorted due to the finite squeezing of the cluster state, which is 
characterized by the convolution of the input Wigner function $W_{\rm in}$ with the momentum space wave functions $G_{\epsilon}\big(x_q \big)$ and $G_{\epsilon} \big( - t x_p \big)$. 

To correct the displacements, we apply $\hat X^{\dag}(m_q)$ and $\hat Z^{\dag}(m_p)$ to the state, which corresponds to performing feedforward in the realistic experiment. 
After the displacement corrections, the Wigner function becomes
\begin{eqnarray}\label{eq:Wigner-7}
&&P(m_1, m_3) W_{\text{out}}(q_{c'}, p_{c'})  \nonumber\\
&=& 
\int \mathrm d x_q \int \mathrm d x_p \frac{2}{|\sin \theta_-|} W_{\text{in}}\bigg[ -\big( x_q +  t q_{c'} \big) \bigg( \frac{ 2 \cos \theta_1 \cos \theta_3}{\sin \theta_-} \bigg) - \bigg( x_p + \frac{ p_{c'}}{t} \bigg) \bigg( \frac{\sin \theta_+}{\sin \theta_-} \bigg),  \nonumber\\
&&
\big( x_q +  t q_{c'} \big) \bigg( \frac{\sin \theta_+}{\sin \theta_-} \bigg) + 
\bigg( x_p + \frac{ p_{c'}}{t} \bigg) \bigg( \frac{2 \sin \theta_1 \sin \theta_3}{\sin \theta_-} \bigg) \bigg] \nonumber\\
&& 
\times G_{1/\epsilon}\bigg( x_p + \frac{p_{c'}}{t} + \frac{m_p}{t} \bigg)
G_{1/\epsilon} \big(q_{c'} + m_q \big) G_{\epsilon}\big(x_q \big)
G_{\epsilon} \big( - t x_p \big). 
\end{eqnarray}

\subsection{Correcting Errors Using Information of Input States}\label{SMEC}

If we are unaware of the input state, the output state is distorted due to the finite squeezing in the cluster state. 
A pure input state would become mixed if one takes into account all measurement outcomes; or it is pure but the wave function is distorted if one post-selects one of the measurement 
outcomes, as can be seen from Eq. \eqref{eq:Wigner-7}. If we know the input state, we can correct the distortions due to the effect of finite squeezing. 

In this paper we only consider Gaussian input states and Gaussian unitaries, therefore the output states are also Gaussian. It is convenient to describe the Gaussian states using
the Wigner function as defined by Eq. \eqref{eq:WignerFunction}.
We assume that the covariance matrix of the input state is $\sigma_{\text{in}}$ and define
\begin{eqnarray}
{\boldsymbol X} = 
 \begin{pmatrix}
 t & 0 \\
 0 & 1/t
 \end{pmatrix}
 \begin{pmatrix}
 q_{c'}  \\
 p_{c'} 
 \end{pmatrix}
 =
 S ( t )
 \begin{pmatrix}
 q_{c'}  \\
 p_{c'} 
 \end{pmatrix},
 ~~~~~~
 {\boldsymbol x} = 
   \begin{pmatrix}
 x_q \\
 x_p 
 \end{pmatrix},
  ~~~~~~
 {\boldsymbol \gamma} = 
  \begin{pmatrix}
 t & 0 \\
 0 & 1/t
 \end{pmatrix}
   \begin{pmatrix}
 m_q \\
 m_p 
 \end{pmatrix}
  =
 S ( t )
 \begin{pmatrix}
 m_q  \\
 m_p
 \end{pmatrix},
\end{eqnarray}
then the input Wigner function in Eq. \eqref{eq:Wigner-7} is 
\begin{eqnarray}
W_{\text{in}} \big[V_a^{-1} ({\boldsymbol X} + {\boldsymbol x}) \big] &=& \frac{1}{2 \pi \sqrt{\text{det}\, \sigma_{\text{in}} }} 
\exp \bigg\{ - \frac{1}{2}({\boldsymbol X}^\top + {\boldsymbol x}^\top) (V_a \, \sigma_{\text{in}} \, V_a^\top)^{-1} ({\boldsymbol X} + {\boldsymbol x}) \bigg\} \nonumber\\
&=&
\frac{1}{2 \pi \sqrt{\text{det}\, \sigma_t }} 
\exp \bigg\{ - \frac{1}{2}({\boldsymbol X}^\top + {\boldsymbol x}^\top) \sigma_t^{-1} ({\boldsymbol X} + {\boldsymbol x}) \bigg\},
\end{eqnarray}
where we have defined the target output covariance matrix as $\sigma_t = V_a \, \sigma_{\text{in}} \, V_a^\top$ and used the fact that 
${\rm det} \, \sigma_{\rm in} = {\rm det} (V_a \, \sigma_{\text{in}} \, V_a^\top )$. 
By using the definition of the Gaussian function, Eq. \eqref{eq:Gaussian},
\begin{eqnarray}
&&G_{1/\epsilon}\bigg( x_p + \frac{p_{c'}}{t} + \frac{m_p}{t} \bigg) G_{1/\epsilon} \big(q_{c'} + m_q \big) G_{\epsilon}\big(x_q \big) G_{\epsilon} \big( - t x_p \big) \nonumber\\
&=&
\frac{1}{\pi^2} \exp \bigg\{-{\boldsymbol x}^\top \Sigma_1 {\boldsymbol x} - ({\boldsymbol X}^\top + {\boldsymbol \gamma}^\top) \Sigma_2 {\boldsymbol x} - {\boldsymbol x}^\top \Sigma_2 ({\boldsymbol X} + {\boldsymbol \gamma})
- ({\boldsymbol X}^\top + {\boldsymbol \gamma}^\top) \Sigma_3 ({\boldsymbol X} + {\boldsymbol \gamma})
\bigg\},
\end{eqnarray}
where 
\begin{eqnarray}\label{eq:Sigma}
\Sigma_1 = 
 \begin{pmatrix}
 1/\epsilon & 0 \\
 0 & \epsilon + t^2/\epsilon
 \end{pmatrix}
 = \frac{1}{\epsilon} \, \mathbb{I}_2, 
 ~~~~~~
 \Sigma_2 = 
 \begin{pmatrix}
 0 & 0 \\
 0 & \epsilon
 \end{pmatrix},
 ~~~~~~
 \Sigma_3 = 
 \begin{pmatrix}
 \epsilon/t^2 & 0 \\
 0 & \epsilon
 \end{pmatrix}. 
\end{eqnarray}
So the exponential of the integrand in \eqref{eq:Wigner-7} is proportional to 
\begin{eqnarray}
&&({\boldsymbol X}^\top + {\boldsymbol x}^\top) \sigma_t^{-1} ({\boldsymbol X} + {\boldsymbol x}) + {\boldsymbol x}^\top (2 \Sigma_1) {\boldsymbol x} + ({\boldsymbol X}^\top + {\boldsymbol \gamma}^\top) (2 \Sigma_2) {\boldsymbol x} 
+ {\boldsymbol x}^\top (2 \Sigma_2) ({\boldsymbol X} + {\boldsymbol \gamma})
+ ({\boldsymbol X}^\top + {\boldsymbol \gamma}^\top) (2 \Sigma_3) ({\boldsymbol X} + {\boldsymbol \gamma}) \nonumber\\
&=& {\boldsymbol x}^\top \big(\sigma_t^{-1}+ 2\Sigma_1 \big) {\boldsymbol x} + {\boldsymbol X}^\top \big(\sigma_t^{-1}+2\Sigma_2 \big) {\boldsymbol x} + {\boldsymbol x}^\top \big(\sigma_t^{-1}+2\Sigma_2 \big) {\boldsymbol X}
+ {\boldsymbol \gamma}^\top (2 \Sigma_2) {\boldsymbol x} + {\boldsymbol x}^\top (2 \Sigma_2) {\boldsymbol \gamma} + {\boldsymbol X}^\top \big(\sigma_t^{-1}+2 \Sigma_3 \big) {\boldsymbol X} \nonumber\\
&&
+ {\boldsymbol X}^\top (2 \Sigma_3) {\boldsymbol \gamma} +  {\boldsymbol \gamma}^\top (2 \Sigma_3) {\boldsymbol X}  + {\boldsymbol \gamma}^\top (2 \Sigma_3) {\boldsymbol \gamma}.
\end{eqnarray}
One can define ${\boldsymbol y} = {\boldsymbol x} - {\boldsymbol x}_0$ with
\begin{eqnarray}
{\boldsymbol x}_0 = - \big({\sigma_t^{-1}+2\Sigma_1} \big)^{-1} \big({\sigma_t^{-1}+2\Sigma_2}  \big) {\boldsymbol X} - \big({\sigma_t^{-1}+2\Sigma_1} \big)^{-1} (2 \Sigma_2) {\boldsymbol \gamma}
\end{eqnarray}
such that the linear terms in ${\bf y}$ disappear, namely, the exponential of the integrand in \eqref{eq:Wigner-7} is proportional to
\begin{eqnarray}
{\boldsymbol y}^\top \big(\sigma_t^{-1}+2 \Sigma_1 \big) {\boldsymbol y} + {\boldsymbol X}^\top \mathcal{A} \, {\boldsymbol X} 
+ {\boldsymbol X}^\top \mathcal{B} \, {\boldsymbol \gamma} + {\boldsymbol \gamma}^\top \mathcal{B}^\top {\boldsymbol X} + {\boldsymbol \gamma}^\top \mathcal{C}\,{\boldsymbol \gamma},
\end{eqnarray}
where we have defined
\begin{eqnarray}
\mathcal{A} &=& \big(\sigma_t^{-1}+2\Sigma_3 \big) - \big(\sigma_t^{-1}+2\Sigma_2 \big) \big(\sigma_t^{-1}+2\Sigma_1 \big)^{-1} {\big(\sigma_t^{-1}+2\Sigma_2 \big)}, \nonumber\\
\mathcal{B} &=& 2\Sigma_3 - \big(\sigma_t^{-1}+2\Sigma_2 \big) \big(\sigma_t^{-1}+2\Sigma_1 \big)^{-1} (2 \Sigma_2), \nonumber\\
\mathcal{C} &=& 2 \Sigma_3 - (2 \Sigma_2) \big(\sigma_t^{-1}+2\Sigma_1 \big)^{-1} (2\Sigma_2). 
\end{eqnarray}
One can further define ${\boldsymbol Y}$ as ${\boldsymbol Y} = {\boldsymbol X} - {\boldsymbol D}$ with
\begin{eqnarray}\label{FDiplacement}
{\boldsymbol D} = - \mathcal{A}^{-1}  \mathcal{B} \, {\boldsymbol \gamma}
\end{eqnarray}
such that the linear terms in ${\boldsymbol Y}$ disappear, then the exponential of the integrand in \eqref{eq:Wigner-7} is proportional to
\begin{eqnarray}
&&{\boldsymbol y}^\top \big(\sigma_t^{-1}+2\Sigma_1 \big) {\bf y} + {\boldsymbol Y}^\top \mathcal{A} \, {\boldsymbol Y} 
+ {\boldsymbol \gamma}^\top \big( \mathcal{C} - \mathcal{B}^\top \mathcal{A}^{-1} \mathcal{B}\big)\,{\boldsymbol \gamma} \nonumber \\
&=& 
\big({\boldsymbol x} - {\boldsymbol x}_0 \big)^\top \big(\sigma_t^{-1}+2\Sigma_1 \big) \big({\boldsymbol x} - {\boldsymbol x}_0 \big) + \big( {\boldsymbol X} - {\boldsymbol D} \big)^\top \mathcal{A} \, \big( {\boldsymbol X} - {\boldsymbol D} \big)
+ {\boldsymbol \gamma}^\top \big( \mathcal{C} - \mathcal{B}^\top \mathcal{A}^{-1} \mathcal{B}\big)\,{\boldsymbol \gamma}. 
\end{eqnarray}
The integration over $\boldsymbol x$ becomes straightforward since it is simply a Gaussian integration. After the integration over $\boldsymbol x$, 
the Wigner function in Eq. \eqref{eq:Wigner-7} becomes 
\begin{eqnarray}\label{eq:Wigner-8}
P(m_1, m_3) W_{\text{out}}(q_{c'}, p_{c'})  
&=& 
 \frac{2}{\pi^2 |\sin \theta_-| \sqrt{\text{det}\,\big(\mathbb{I}_2+2\sigma_t \Sigma_1 \big)} } 
\exp \bigg\{ - \frac{1}{2}\big( {\boldsymbol X} - {\boldsymbol D} \big)^\top \mathcal{A} \, \big( {\boldsymbol X} - {\boldsymbol D} \big)
- \frac{1}{2} {\boldsymbol \gamma}^\top \big( \mathcal{C} - \mathcal{B}^\top \mathcal{A}^{-1} \mathcal{B}\big)\,{\boldsymbol \gamma} \bigg\}. \nonumber\\
\end{eqnarray}
We can further apply displacements to $q_{c'}$ and $p_{c'}$, namely, ${\boldsymbol X} \rightarrow {\boldsymbol X} + {\boldsymbol D}$. The displacement 
vector ${\boldsymbol D}$ is given by Eq. \eqref{FDiplacement}, and is dependent on the input covariance matrix $\sigma_{\text{in}}$, the unitary $\hat V_a$ and
the homodyne measurement outcomes. After this displacement and averaging over all measurement outcomes, the output Wigner function becomes
\begin{eqnarray}\label{eq:Wigner-output}
W_{\text{out}}(q_{c'}, p_{c'}) 
&=&
\int \mathrm d m_1 \int \mathrm d m_3  ~ P(m_1, m_3) W_{\text{out}}(q_{c'}, p_{c'})  
=
\frac{1}{2\pi \sqrt{\text{det}\, \sigma}}  \exp \bigg\{ - \frac{1}{2}{\boldsymbol \xi}_{c'}^\top \sigma^{-1} \, {\boldsymbol \xi}_{c'} \bigg\},
\end{eqnarray}
where 
$\sigma = S \big(\frac{1}{t} \big) \mathcal{A}^{-1} S^\top \big(\frac{1}{t} \big)$ is the actual covariance matrix after performing the additional displacement correction 
by using the information of the input state. $\sigma$ is still different from the target covariance matrix $\sigma_t$. 
\begin{eqnarray}
S^{-\top} (t) \, \sigma^{-1} S^{-1} (t) = \mathcal{A} 
=
\sigma_t^{-1} \big( \mathbb{I}_2 + \Delta_{\sigma} \big),
\end{eqnarray}
where 
\begin{eqnarray}
\Delta_{\sigma} \equiv 2\sigma_t \Sigma_3 - \big( \mathbb{I}_2 + 2\sigma_t \Sigma_2 \big) \big( \mathbb{I}_2 + 2 \Sigma_1 \sigma_t  \big)^{-1} \big( \mathbb{I}_2 + 2\sigma_t \Sigma_2 \big)
\end{eqnarray}
characterizes the difference between the target covariance matrix and the actual covariance matrix. It can be shown from Eq. \eqref{eq:Sigma} 
that $\Delta_{\sigma} \rightarrow 0$ when $\epsilon \rightarrow 0$.
The target covariance matrix can be written in terms of the actual output covariance matrix $\sigma$ as
\begin{eqnarray}
\sigma_t = \big( \mathbb{I}_2 + \Delta_{\sigma} \big) S(t) \sigma S^\top(t). 
\end{eqnarray}

\section{Two-mode Unitary Errors Due to Finite Squeezing}\label{TMEC}

In this appendix, we focus on implementing a two-mode unitary by performing four homodyne measurements, and correcting the errors due to the
finite squeezing in the CV cluster state. The scheme of implementing a measurement-based two-mode unitary is shown in Fig. \ref{fig:TMU}. By comparing it with Fig. \ref{fig:SMU},
we note that each green box in Fig. \ref{fig:TMU} basically implements a single-mode unitary. A two-mode unitary is achieved by sandwiching two single-mode unitaries with
two $50:50$ beam splitters: $B(\pi/4)$ and $B(-\pi/4)$. Based on this observation, we can proceed with the help of the analysis in Appendix \ref{SMError}. 

To simplify the analysis, we assume that the Wigner function after the first beam splitter $B(\pi/4)$ is $W_2(q_a, p_a, q_b, p_b)$, which is related to the input Wigner function
$W_{\text{in}}$ via
\begin{eqnarray}
W_2(q_a, p_a, q_b, p_b) = W_{\text{in}} \bigg[\frac{1}{\sqrt{2}} (q_a + q_b), \frac{1}{\sqrt{2}} (p_a + p_b), \frac{1}{\sqrt{2}} (-q_a + q_b), \frac{1}{\sqrt{2}} (-p_a + p_b) \bigg], 
\end{eqnarray}
Following $W_2(q_a, p_a, q_b, p_b)$ are two independent measurement-based single-mode unitaries, which we label as $\hat V_a$ and $\hat V_b$, respectively.
We have analyzed $\hat V_a$ in Appendix \ref{SMError}. The analysis of $\hat V_b$ can be done similarly and results can be obtained by replacing ``1" by ``2"
and ``3" by ``4". To avoid confusions we assume all notations for the unitary $\hat V_a$ are the same as those in Appendix \ref{SMError} 
and introduce different notations for the second unitary $\hat V_b$. Define $\varphi_{\pm} = \theta_2 \pm \theta_4$ and
\begin{eqnarray}
\bar m_q &=& \frac{\sqrt{2} \big( m_2 \sin \theta_4 + m_4 \sin \theta_2 \big)}{t \sin \varphi_-}, \nonumber\\
\bar m_p &=& - \frac{\sqrt{2} t \big( m_2 \cos \theta_4 + m_4 \cos \theta_2 \big)}{\sin \varphi_-}. 
\end{eqnarray}
The symplectic transformation corresponding to the unitary $\hat V_b$ is 
\begin{eqnarray}
V_b(\theta_2, \theta_4) &=&  R_b \big(\varphi_+/2  \big) S_b \big[ \tan (\varphi_-/2)  \big] R_b \big(\varphi_+/2  \big) R_b(\pi)
\end{eqnarray}
The Wigner function before the beam splitter $B(-\pi/4)$ (after displacement corrections of each single-mode unitary) is
\begin{eqnarray}\label{eq:two-Wigner-1}
&&P(m_1, m_2, m_3, m_4) W_{\text{out}}(q_{c'}, p_{c'}, q_{d'}, p_{d'})  \nonumber\\
&=& 
\int \mathrm d x_q \int \mathrm d x_p  \int \mathrm d y_q \int \mathrm d y_p \frac{4}{|\sin \theta_- \, \sin \varphi_-|} 
W_2\bigg[ -\big( x_q +  t q_{c'} \big) \bigg( \frac{ 2 \cos \theta_1 \cos \theta_3}{\sin \theta_-} \bigg) - \bigg( x_p + \frac{ p_{c'}}{t} \bigg) \bigg( \frac{\sin \theta_+}{\sin \theta_-} \bigg),  \nonumber\\
&&
\big( x_q +  t q_{c'} \big) \bigg( \frac{\sin \theta_+}{\sin \theta_-} \bigg) + \bigg( x_p + \frac{ p_{c'}}{t} \bigg) \bigg( \frac{2 \sin \theta_1 \sin \theta_3}{\sin \theta_-} \bigg),
-\big( y_q +  t q_{d'} \big) \bigg( \frac{ 2 \cos \theta_2 \cos \theta_4}{\sin \varphi_-} \bigg) - \bigg( y_p + \frac{ p_{d'}}{t} \bigg) \bigg( \frac{\sin \varphi_+}{\sin \varphi_-} \bigg),  \nonumber\\
&&
\big( y_q +  t q_{d'} \big) \bigg( \frac{\sin \varphi_+}{\sin \varphi_-} \bigg) + \bigg( y_p + \frac{ p_{d'}}{t} \bigg) \bigg( \frac{2 \sin \theta_2 \sin \theta_4}{\sin \varphi_-} \bigg) \bigg] 
\nonumber\\
&& 
\times G_{1/\epsilon}\bigg( x_p + \frac{p_{c'}}{t} + \frac{m_p}{t} \bigg)
G_{1/\epsilon} \big(q_{c'} + m_q \big) G_{\epsilon}\big(x_q \big) G_{\epsilon} \big( - t x_p \big)
\nonumber\\
&& 
\times G_{1/\epsilon}\bigg( y_p + \frac{p_{d'}}{t} + \frac{\bar m_p}{t} \bigg)
G_{1/\epsilon} \big(q_{d'} + \bar m_q \big) G_{\epsilon}\big(y_q \big) G_{\epsilon} \big( - t y_p \big), 
\end{eqnarray}
where $P(m_1, m_2, m_3, m_4)$ is the probability of registering measurement outcomes $m_1, m_2, m_3$ and $m_4$ simultaneously. 
We further define 
\begin{eqnarray}
M_q = \frac{1}{\sqrt{2}} (m_q + \bar m_q), ~~~~ \tilde M_q = \frac{1}{\sqrt{2}} (-m_q + \bar m_q), ~~~~ 
M_p = \frac{1}{\sqrt{2}} (m_p + \bar m_p), ~~~~ \tilde M_p = \frac{1}{\sqrt{2}} (-m_p + \bar m_p).  
\end{eqnarray}
After the last beam spitter $B(-\pi/4)$, the Wigner function becomes
\begin{eqnarray}\label{eq:two-Wigner-2}
&&P(m_1, m_2, m_3, m_4) W_{\text{out}}(q_{c'}, p_{c'}, q_{d'}, p_{d'})  \nonumber\\
&=& 
\int \mathrm d x_q \int \mathrm d x_p  \int \mathrm d y_q \int \mathrm d y_p \frac{4}{|\sin \theta_- \, \sin \varphi_-|} 
W_{\text{in}} \bigg[ B \bigg(-\frac{\pi}{4} \bigg) \big(V_a \oplus V_b \big)^{-1} B \bigg(\frac{\pi}{4} \bigg) \big( \tilde{\boldsymbol X} + \tilde{\boldsymbol x}^{\prime} \big) \bigg] 
\nonumber\\
&& 
\times G_{1/\epsilon}\bigg[ x_p + \frac{(p_{c'}+M_p) - (p_{d'} + \tilde M_p) }{\sqrt{2} \, t} \bigg]
G_{1/\epsilon} \bigg[\frac{(q_{c'}+M_q) - (q_{d'} + \tilde M_q) }{\sqrt{2}} \bigg] G_{\epsilon}\big(x_q \big) G_{\epsilon} \big( - t x_p \big)
\nonumber\\
&& 
\times G_{1/\epsilon}\bigg[ y_p + \frac{(p_{c'}+M_p) + (p_{d'} + \tilde M_p) }{\sqrt{2} \, t} \bigg]
G_{1/\epsilon} \bigg[\frac{(q_{c'}+M_q) + (q_{d'} + \tilde M_q) }{\sqrt{2}} \bigg] G_{\epsilon}\big(y_q \big) G_{\epsilon} \big( - t y_p \big),
\end{eqnarray}
where we have defined $\tilde{\boldsymbol x}^{\prime} = B(-\pi/4) \tilde {\boldsymbol x}$ with $\tilde {\boldsymbol x} = (x_q, x_p, y_q, y_p)^\top$, and
\begin{eqnarray}
\tilde{\boldsymbol X} = 
\begin{pmatrix}
 t & 0 & 0 & 0 \\
 0 & 1/t & 0 & 0 \\
 0 & 0& t & 0 \\
 0 & 0 & 0& 1/t 
 \end{pmatrix}
 \begin{pmatrix}
 q_{c'}  \\
 p_{c'} \\
 q_{d'} \\
 p_{d'}
 \end{pmatrix}
 \equiv
 \tilde S(t)
 \begin{pmatrix}
 q_{c'}  \\
 p_{c'} \\
 q_{d'} \\
 p_{d'}
 \end{pmatrix}. 
\end{eqnarray}
Note that we also used the fact that $B(\pi/4)$ commutes with $\tilde S(t)$.
By using the definition of the Gaussian function, Eq. \eqref{eq:Gaussian},
\begin{eqnarray}
&&G_{1/\epsilon}\bigg[ x_p + \frac{(p_{c'}+M_p) - (p_{d'} + \tilde M_p) }{\sqrt{2} \, t} \bigg]
G_{1/\epsilon} \bigg[\frac{(q_{c'}+M_q) - (q_{d'} + \tilde M_q) }{\sqrt{2}} \bigg] G_{\epsilon}\big(x_q \big) G_{\epsilon} \big( - t x_p \big)
\nonumber\\
&& 
\times G_{1/\epsilon}\bigg[ y_p + \frac{(p_{c'}+M_p) + (p_{d'} + \tilde M_p) }{\sqrt{2} \, t} \bigg]
G_{1/\epsilon} \bigg[\frac{(q_{c'}+M_q) + (q_{d'} + \tilde M_q) }{\sqrt{2}} \bigg] G_{\epsilon}\big(y_q \big) G_{\epsilon} \big( - t y_p \big) \nonumber\\
&=&
\frac{1}{\pi^4} \exp \bigg\{- \tilde {\boldsymbol x}^{\prime \top} \tilde \Sigma_1 \tilde {\boldsymbol x}^{\prime} - (\tilde {\boldsymbol X}^\top + \tilde {\boldsymbol \gamma}^\top) {\tilde \Sigma}_2^\top \tilde {\boldsymbol x}^{\prime} - \tilde {\boldsymbol x}^{\prime \top} \tilde \Sigma_2 (\tilde {\boldsymbol X} + \tilde {\boldsymbol \gamma})
- (\tilde {\boldsymbol X}^\top + \tilde {\boldsymbol \gamma}^\top) \tilde \Sigma_3 (\tilde {\boldsymbol X} + \tilde {\boldsymbol \gamma})
\bigg\},
\end{eqnarray}
where
\begin{eqnarray}
\tilde{\boldsymbol \gamma} = 
\begin{pmatrix}
 t & 0 & 0 & 0 \\
 0 & 1/t & 0 & 0 \\
 0 & 0& t & 0 \\
 0 & 0 & 0& 1/t 
 \end{pmatrix}
 \begin{pmatrix}
 M_q  \\
 M_p \\
 \tilde M_q \\
 \tilde M_p
 \end{pmatrix}
\end{eqnarray}
and
\begin{eqnarray}\label{eq:Sigma-two}
\tilde \Sigma_1 &=& B^\top \bigg(\frac{\pi}{4} \bigg) \big( \Sigma_1 \oplus \Sigma_1 \big) B \bigg(\frac{\pi}{4} \bigg) = \frac{1}{\epsilon} \, \mathbb{I}_4, \nonumber\\
 ~~~~~~
\tilde \Sigma_2 &=& B^\top \bigg(\frac{\pi}{4} \bigg) \big( \Sigma_2 \oplus \Sigma_2 \big) B \bigg(\frac{\pi}{4} \bigg) 
 =
 \begin{pmatrix}
 \Sigma_2 & 0 \\
 0 & \Sigma_2 
 \end{pmatrix},
 \nonumber\\
 \tilde \Sigma_3 &=& B^\top \bigg(\frac{\pi}{4} \bigg) \big( \Sigma_3 \oplus \Sigma_3 \big) B \bigg(\frac{\pi}{4} \bigg) =
 \begin{pmatrix}
 \Sigma_3 & 0 \\
 0 & \Sigma_3 
 \end{pmatrix}. 
\end{eqnarray}
So the exponential of the integrand in \eqref{eq:two-Wigner-2} is proportional to 
\begin{eqnarray}
&&(\tilde {\boldsymbol X}^\top + \tilde {\boldsymbol x}^{\prime \top}) \sigma_t^{-1} ( \tilde {\boldsymbol X} + \tilde {\boldsymbol x}^{\prime}) + \tilde {\boldsymbol x}^{\prime \top} (2 \tilde \Sigma_1) \tilde {\boldsymbol x}^{\prime} + (\tilde {\boldsymbol X}^\top + \tilde {\boldsymbol \gamma}^\top) (2 \tilde \Sigma_2) \tilde {\boldsymbol x}^{\prime} + \tilde {\boldsymbol x}^{\prime \top} (2 \tilde \Sigma_2) (\tilde {\boldsymbol X} + \tilde {\boldsymbol \gamma})
+ (\tilde {\boldsymbol X}^\top + \tilde {\boldsymbol \gamma}^\top) (2 \tilde \Sigma_3) (\tilde {\boldsymbol X} + \tilde {\boldsymbol \gamma}) \nonumber\\
&=& \tilde {\boldsymbol x}^{\prime \top} \big(\sigma_t^{-1}+ 2 \tilde \Sigma_1 \big) \tilde {\boldsymbol x}^{\prime} + \tilde {\boldsymbol X}^\top \big(\sigma_t^{-1}+2 \tilde \Sigma_2 \big) \tilde{\boldsymbol x}^{\prime} + \tilde {\boldsymbol x}^{\prime \top} \big(\sigma_t^{-1}+2 \tilde \Sigma_2 \big) \tilde {\boldsymbol X}
+ \tilde {\boldsymbol \gamma}^\top (2 \tilde \Sigma_2) \tilde {\boldsymbol x}^{\prime} + \tilde {\boldsymbol x}^{\prime \top} (2 \tilde \Sigma_2) \tilde {\boldsymbol \gamma} + \tilde {\boldsymbol X}^\top \big(\sigma_t^{-1}+2 \tilde \Sigma_3 \big) \tilde {\boldsymbol X} \nonumber\\
&&
+ \tilde {\boldsymbol X}^\top (2 \tilde \Sigma_3) \tilde {\boldsymbol \gamma} +  \tilde {\boldsymbol \gamma}^\top (2 \tilde \Sigma_3) \tilde {\boldsymbol X}  + \tilde {\boldsymbol \gamma}^\top (2 \tilde \Sigma_3) \tilde {\boldsymbol \gamma}.
\end{eqnarray}
The target covariance matrix $\sigma_t$ is 
\begin{eqnarray}
\sigma_t = \bigg[ B \bigg(-\frac{\pi}{4} \bigg) \big(V_a \oplus V_b \big) B \bigg(\frac{\pi}{4} \bigg) \bigg] \sigma_{\text{in}} 
\bigg[ B \bigg(-\frac{\pi}{4} \bigg) \big(V_a \oplus V_b \big)^\top B \bigg(\frac{\pi}{4} \bigg) \bigg]
\equiv V \sigma_{\text{in}} V^\top.
\end{eqnarray}
By introducing $\tilde {\boldsymbol y} = \tilde {\boldsymbol x}^{\prime} - \tilde {\boldsymbol x}_0^{\prime}$  with
%
\begin{eqnarray}
\tilde {\boldsymbol x}_0^{\prime} = - \big({\sigma_t^{-1}+2 \tilde \Sigma_1} \big)^{-1} \big({\sigma_t^{-1}+2\tilde \Sigma_2}  \big) \tilde {\boldsymbol X} - \big({\sigma_t^{-1}+2 \tilde \Sigma_1} \big)^{-1} (2 \tilde \Sigma_2) \tilde {\boldsymbol \gamma}
\end{eqnarray}
such that the linear terms in $\tilde {\boldsymbol y}$ disappear, namely, the exponential of the integrand in \eqref{eq:two-Wigner-2} is proportional to
\begin{eqnarray}
\tilde {\bf y}^\top \big(\sigma_t^{-1}+2 \tilde \Sigma_1 \big) \tilde {\bf y} + \tilde {\bf X}^\top \tilde{\mathcal{A}} \, \tilde {\bf X} 
+ \tilde {\bf X}^\top \tilde{\mathcal{B}} \, \tilde {\boldsymbol \gamma} + \tilde {\boldsymbol \gamma}^\top \tilde{\mathcal{B}}^\top \tilde {\bf X} + \tilde {\boldsymbol \gamma}^\top \tilde{\mathcal{C}}\,\tilde{\boldsymbol \gamma},
\end{eqnarray}
where
\begin{eqnarray}
\tilde{\mathcal{A}} &=& \big(\sigma_t^{-1}+2 \tilde \Sigma_3 \big) - \big(\sigma_t^{-1}+2 \tilde \Sigma_2 \big) \big(\sigma_t^{-1}+2 \tilde \Sigma_1 \big)^{-1} {\big(\sigma_t^{-1}+2 \tilde \Sigma_2 \big)}, \nonumber\\
\tilde{\mathcal{B}} &=& 2 \tilde \Sigma_3 - \big(\sigma_t^{-1}+2 \tilde \Sigma_2 \big) \big(\sigma_t^{-1}+2 \tilde \Sigma_1 \big)^{-1} (2 \tilde \Sigma_2), \nonumber\\
\tilde{\mathcal{C}} &=& 2 \tilde \Sigma_3 - (2 \tilde \Sigma_2) \big(\sigma_t^{-1}+2 \tilde \Sigma_1 \big)^{-1} (2 \tilde \Sigma_2). 
\end{eqnarray}
One can further define $\tilde {\boldsymbol Y} = \tilde {\boldsymbol X} - \tilde {\boldsymbol D}$ with
\begin{eqnarray}\label{FDiplacement-two}
\tilde {\boldsymbol D} = - \tilde{\mathcal{A}}^{-1}  \tilde{\mathcal{B}} \, \tilde {\boldsymbol \gamma}
\end{eqnarray}
such that the linear terms in $\tilde {\boldsymbol Y}$ disappear, then the exponential of the integrand in \eqref{eq:two-Wigner-2} is proportional to
\begin{eqnarray}
&&\tilde {\boldsymbol y}^\top \big(\sigma_t^{-1}+2 \tilde \Sigma_1 \big) \tilde {\boldsymbol y} + \tilde {\boldsymbol Y}^\top \tilde{\mathcal{A}} \, \tilde {\boldsymbol Y} 
+ \tilde {\boldsymbol \gamma}^\top \big( \tilde{\mathcal{C}} - \tilde{\mathcal{B}}^\top \tilde{\mathcal{A}}^{-1} \tilde{\mathcal{B}} \big)\,\tilde {\boldsymbol \gamma} \nonumber \\
&=& 
\big(\tilde {\boldsymbol x}^{\prime} - \tilde {\boldsymbol x}_0^{\prime} \big)^\top \big(\sigma_t^{-1}+2 \tilde \Sigma_1 \big) \big(\tilde {\boldsymbol x}^{\prime} - \tilde {\boldsymbol x}_0^{\prime} \big) + \big( \tilde {\boldsymbol X} - \tilde {\boldsymbol D} \big)^\top \tilde{\mathcal{A}} \,\big( \tilde {\boldsymbol X} - \tilde {\boldsymbol D} \big)
+ \tilde {\boldsymbol \gamma}^\top \big( \tilde{\mathcal{C}} - \tilde{\mathcal{B}}^\top \tilde{\mathcal{A}}^{-1} \tilde{\mathcal{B}} \big)\,\tilde {\boldsymbol \gamma}. 
\end{eqnarray}
After performing the integration over $\tilde {\boldsymbol x}^{\prime}$, which is simply a Gaussian integration, the Wigner function \eqref{eq:two-Wigner-2} becomes
\begin{eqnarray}\label{eq:two-Wigner}
&&P(m_1, m_2, m_3, m_4) W_{\text{out}}(q_{c'}, p_{c'}, q_{d'}, p_{d'})  \nonumber\\
&=&
\frac{4}{\pi^4 |\sin \theta_- \, \sin \varphi_-| \sqrt{\text{det} \, (\mathbb{I}_4 + 2\sigma_t \tilde \Sigma_1)} } \exp \bigg\{ -\frac{1}{2}\big( \tilde {\boldsymbol X} - \tilde {\boldsymbol D} \big)^\top \tilde{\mathcal{A}} \,\big( \tilde {\boldsymbol X} - \tilde {\boldsymbol D} \big)
- \frac{1}{2}\tilde {\boldsymbol \gamma}^\top \big( \tilde{\mathcal{C}} - \tilde{\mathcal{B}}^\top \tilde{\mathcal{A}}^{-1} \tilde{\mathcal{B}} \big)\,\tilde {\boldsymbol \gamma} \bigg\}. 
\end{eqnarray}
We can further apply displacements to $q_{c'}$, $p_{c'}$, $q_{d'}$ and $p_{d'}$, specifically, $\tilde {\boldsymbol X} \rightarrow \tilde {\boldsymbol X} + \tilde {\boldsymbol D}$. 
The displacement $\tilde {\boldsymbol D}$ is given by Eq. \eqref{FDiplacement-two}, and
is dependent on the input covariance matrix $\sigma_{\text{in}}$, the unitary $\hat V$ and the homodyne measurement outcomes. 
After this displacement and averaging over all measurement outcomes, the actual output Wigner function becomes
\begin{eqnarray}\label{eq:Wigner-output}
W(q_{c'}, p_{c'}, q_{d'}, p_{d'}) &=& 
\int \mathrm d m_1 \int \mathrm d m_2 \int \mathrm d m_3 \int \mathrm d m_4 ~ P(m_1, m_2, m_3, m_4) W_{\text{out}}(q_{c'}, p_{c'}, q_{d'}, p_{d'})  \nonumber\\
&=&
\frac{1}{(2\pi)^2 \sqrt{\text{det}\, \sigma_2}}  \exp \bigg\{ - \frac{1}{2}{\boldsymbol \xi}^{\prime \top} \sigma_2^{-1} \, {\boldsymbol \xi}^{\prime} \bigg\},
\end{eqnarray}
where 
$\sigma_2 = \tilde S \big(\frac{1}{t} \big) \tilde{\mathcal{A}}^{-1} \tilde S^\top \big(\frac{1}{t} \big)$ is the covariance matrix after performing the additional displacement correction by using the information of the input state. $\sigma_2$ is still different from the target covariance matrix $\sigma_t$. 
\begin{eqnarray}
\tilde S^{-\top} (t) \, \sigma_2^{-1} \tilde S^{-1} (t) = \tilde{\mathcal{A}}
&=&
\sigma_t^{-1} \big( \mathbb{I}_4 + \tilde \Delta_{\sigma} \big),
\end{eqnarray}
where 
\begin{eqnarray}\label{Error:twomode}
\tilde \Delta_{\sigma} \equiv 2 \sigma_t \tilde \Sigma_3 - \big( \mathbb{I}_4 + 2 \sigma_t \tilde \Sigma_2 \big) \big( \mathbb{I}_4 + 2 \tilde \Sigma_1 \sigma_t  \big)^{-1} \big( \mathbb{I}_4 + 2 \sigma_t \tilde \Sigma_2 \big)
\end{eqnarray}
characterizes the difference between the target covariance matrix and the actual covariance matrix. It can be shown from Eq. \eqref{eq:Sigma-two} 
that $\tilde \Delta_{\sigma} \rightarrow 0$ when $\epsilon \rightarrow 0$.
The target covariance matrix can be written as
\begin{eqnarray}
\sigma_t = \big( \mathbb{I}_4 + \tilde \Delta_{\sigma} \big) \tilde S(t) \sigma_2 \tilde S^\top(t). 
\end{eqnarray}

\end{widetext}


\end{document}